\documentclass[sigconf,balance=false]{acmart}
\usepackage{popets}
\usepackage[normalem]{ulem}

\setcopyright{popets}
\copyrightyear{YYYY}
\acmYear{YYYY}
\acmVolume{YYYY}
\acmNumber{X}
\acmDOI{XXXXXXX.XXXXXXX}
\acmISBN{}
\acmConference{Proceedings on Privacy Enhancing Technologies}
\newif\ifcomments
\commentstrue

\ifcomments
\newcommand{\ts}[1]{\textcolor{blue}{\textbf{TS:} #1}}
\newcommand{\jt}[1]{\textcolor{purple}{\textbf{JT:} #1}}
\newcommand{\sh}[1]{\textcolor{olive}{\textbf{SH:} #1}}
\newcommand{\dan}[1]{\textcolor{blue}{\textbf{Dan:} #1}}
\newcommand{\new}[1]{#1}
\newcommand{\newnew}[1]{#1}
\newcommand{\remove}[1]{\textcolor{red}{\sout{#1}}}
\else
\newcommand{\ts}[1]{}
\newcommand{\jt}[1]{}
\newcommand{\sh}[1]{}
\newcommand{\dan}[1]{}
\newcommand{\new}[1]{#1}
\newcommand{\remove}[1]{}
\fi

\begin{document}

\title{A False Sense of Privacy: Towards a Reliable Evaluation Methodology for the Anonymization of Biometric Data}


\author{Simon Hanisch}
\affiliation{%
  \institution{Center for Tactile Internet (CeTI), Technical University Dresden}
  \city{Dresden}
  \state{}
  \country{Germany}}
\email{simon.hanisch@tu-dresden.de}

\author{Julian Todt}
\affiliation{%
  \institution{KASTEL,\\Karlsruhe Institute of Technology}
  \city{Karlsruhe}
  \state{}
  \country{Germany}}
\email{julian.todt@kit.edu}

\author{Jose Patino}
\affiliation{%
  \institution{Cerence$^{\ast}$}
  \city{Burlington}
  \state{}
  \country{United States}
}
\email{jose.patino@cerence.com}
\thanks{$^{\ast}$Work done while at EURECOM}

\author{Nicholas Evans}
\affiliation{%
  \institution{Digital Security Department, EURECOM}
  \city{Biot - Sophia Antipolis}
  \state{}
  \country{France}
}
\email{evans@eurecom.fr}

\author{Thorsten Strufe}
\affiliation{%
  \institution{KASTEL,\\Karlsruhe Institute of Technology}
  \city{Karlsruhe}
  \state{}
  \country{Germany}}
\email{thorsten.strufe@kit.edu}


\renewcommand{\shortauthors}{Hanisch et al.}

\begin{abstract}
Biometric data contains distinctive human traits such as facial features or gait patterns. The use of biometric data permits an individuation so exact that the data is utilized effectively in identification and authentication systems. But for this same reason, privacy protections become indispensably necessary.

Privacy protection is extensively afforded by the technique of anonymization. Anonymization techniques protect sensitive personal data from biometrics by obfuscating or removing information that allows linking records to the generating individuals, to achieve high levels of anonymity. However, our understanding and possibility to develop effective  anonymization relies, in equal parts, on the effectiveness of the methods employed to evaluate anonymization performance.

In this paper, we assess the state-of-the-art methods used to evaluate the performance of anonymization techniques for facial images and for gait patterns. We demonstrate that the state-of-the-art evaluation methods have serious and frequent shortcomings. In particular, we find that the underlying assumptions of the state-of-the-art are quite unwarranted. State-of-the-art methods generally assume a difficult recognition scenario and thus a weak adversary. However, that assumption causes state-of-the-art evaluations to grossly overestimate the performance of the anonymization. Therefore, we propose a strong adversary which is aware of the anonymization in place. This adversary model implements an appropriate measure of anonymization performance. We improve the selection process for the evaluation dataset, and we reduce the numbers of identities contained in the dataset while ensuring that these identities remain easily distinguishable from one another. Our novel evaluation methodology surpasses the state-of-the-art because we measure worst-case performance and so deliver a highly reliable evaluation of biometric anonymization techniques.
\end{abstract}

\keywords{privacy, biometric, methodology, anonymization, evaluation}

\maketitle

\section{Introduction}
Biometric data is rich in sensitive personal information that can be used to identify individuals and infer private attributes. \new{Usage e}xamples of biometric data are face recognition~\cite{deng_arcface_2019}, gait recognition~\cite{Wan_Gait_Survey_2018}, inference of medical conditions~\cite{Koktas_Medical_2006}, and inference of character traits based on eye motion~\cite{kroger2020does}. \new{Thus, the utility provided by biometric data is undeniable. Users upload images to social media, share videos with friends, and use online services for health tracking. However, users also care about personal privacy, and in this connection, the use of biometric data poses a real and serious threat. Biometric data allows to draw conclusions about the sensitive personal information of users without their explicit consent.}

\new{I}n order to protect the privacy of the individuals whose biometrics are captured, techniques have been developed that perturb biometric data and so obfuscate or remove their sensitive personal information, or information that may allow for linking them to their generating individuals. \new{Besides protection, these techniques are also designed to maintain the practical utility of the biometric use case. In short, privacy-protecting techniques make the trade-off between utility and protection.} The representative technique \new{in identity protection} is anonymization. 

\new{Identity protection through anonymization ultimately depends on reliable evaluation of anonymization performance.} \new{A} reliable evaluation methodology for anonymization will accurately assess the level of protection afforded against \new{identification} \new{by} the anonymized data.
\new{Reliable evaluation begins at the} assumptions made about an attacker\new{. These assumptions} must be robust, because otherwise the evaluation methodology will likely deliver grossly inaccurate estimates of anonymization performance. The result will be a false sense of privacy, and the consequence will be the erosion of user trust. Moreover, any inaccuracy or even error in an evaluation methodology will detrimentally affect advances in the research. Flaws in the methodology may feed into future research and thus hinder or even arrest the development of advanced anonymization techniques. The upshot is this: Only when a biometric anonymization technique has been convincingly evaluated can researchers improve on existing techniques or provide secure applications to users.

In this paper, we assess the state-of-the-art evaluation methodology for the anonymization of biometric data. In particular, we assess the \new{evaluative }methods for face anonymization and gait anonymization. Our choice for face anonymization is based on the fact that there are many widely-employed techniques in application. On the other hand, we have chosen gait anonymization precisely because the techniques here are fewer and relatively seldom in application. Moreover, the 3D time-series data in gait anonymization allow us to push the envelope of the simpler two-dimensional space of the face image. \new{We acknowledge} that the comprehensive evaluation of any anonymization technique is only possible when utility is also taken into consideration. However, for this paper, we have narrowed our scope to the improvement of the methods evaluating \new{privacy protection of} anonymization only.


\new{Our assessment of the state-of-the-art in evaluation for the anonymization of biometric data shows that these methods often fail at convincingly evaluating the performance of the privacy protection.}

\new{The state-of-the-art methods have been uncritically adopted from the evaluation methodology for biometric recognition. In biometric recognition, the problems employ many identities and difficult biometric samples (e.g., profile photos or nearly indistinguishable identities). In anonymization, on the other hand, a difficult problem has a small number of identities which are very diverse, thus making the identities easier to differentiate but more difficult to anonymize.}

\new{Furthermore, the state-of-the-art methods rely on weak adversary models. These methods assume that the attacker is unaware of the anonymization mechanisms in place. For example, a method will use pre-trained recognition models which perform well on clear data. However, such models prove incapable of adapting to data modifications performed by an anonymization technique. Consider this straightforward scenario: An anonymization technique for a face image performs consistently the same block permutation. This anonymization can easily be removed with the inverse permutation. However, the permutation will go unnoticed by a recognition model pre-trained on the clear data. Moreover, if only a single recognition is used, then that will jeopardize the reliability of the evaluation. Although a given anonymization technique may successfully degrade the performance of one recognition system, \textit{other} systems classifying \textit{other} feature vectors may be more robust or even largely unaffected. However, the use of just a single recognition system is the norm among state-of-the-art evaluation methods.}

\new{We conclude that improvements to the state-of-the-art evaluation for the anonymization of biometric data will also improve the privacy protection offered by anonymization techniques. To this end, we offer specific recommendations for the improvement of state-of-the-art evaluation methods. We draw special attention to just three of our recommendations here. First, we recommend training and (where in use) pre-training recognition systems on anonymized data. Second, we recommend considering multiple and different recognition systems. Third, we recommend choosing smaller datasets in an informed manner.}

The contributions of our paper are as follows:
\begin{itemize}
\item We assess the current state-of-the-art evaluation methodology for biometric data anonymization and point to fatal flaws in the evaluation \new{methodology}. 
\item We update the state-of-the-art evaluation methodology. Our methodological improvements involve (1) retraining the recognition system on anonymized data, (2) using multiple recognition systems to evaluate the anonymization, and (3) generating evaluation datasets that are challenging to \new{anonymize} and consequently reliable for the evaluation of the \new{anonymization} performance.
\item We test our methodological improvements on the biometric traits face and gait \new{with extensive experimentation}. Our evidence supports the conclusion that our improved methodology delivers reliable evaluations of biometric data anonymization.
\end{itemize}

Here we outline the organization of our paper. In Section~\ref{sec:related_work}, we introduce the related work and continue in Section~\ref{sec:background} by giving the background. In Section~\ref{sec:evaluation_methodology}, we present our improved methodology for the evaluation of anonymization techniques. In Section~\ref{sec:evaluation}, we setup our experiments on both face and gait recognition, and in Section~\ref{sec:results} we analyze our results to show how our methodology improves upon the state-of-the-art. In Section~\ref{sec:future_work}, we discuss our findings and explore the future work in this important area of privacy research. And in Section~\ref{sec:conclusion}, we draw our conclusions.

\section{Related Work}\label{sec:related_work}
Biometric recognition spans dozens of biometric traits and hundreds of techniques, but the methodology for evaluating the performance of these techniques has been assessed by only a very few works.

Goga et al.~\cite{goga_reliability_2015} assess the methodology for evaluating matching techniques of profiles from different social media platforms. They find that evaluation commonly overestimates the performance of the approaches by using an unrealistic methodology. Granger and Gorodnischy~\cite{granger_evaluation_2014} describe the methodology that should be applied to evaluate the performance of biometric recognition for video surveillance applications. For the evaluation of stylometric authorship attribution, Stolerman et al.~\cite{stolerman_classify_nodate} make the case that an open-set model should be applied since in a realistic scenario the actual author might not be on the suspect list. Brennan et al.~\cite{brennan_adversarial_2012} propose adding attacks to the methodology of stylometry evaluation because most methods cannot defend against attacks. \new{These investigations of the evaluation methodology in different fields have shown that wrong assumptions lead to an overestimation of performance. In the case of anonymization, overestimation of performance may give users false assurances of privacy because, in fact, their identities are actually left unprotected. In this paper we similarly look at a current evaluation methodology, highlight issues and propose solutions.}


Template protection is a very specific kind of privacy protection because it removes all possible attribute inferences while still allowing identity verification. One specialized evaluation methodology~\cite{rane_standardization_2014} for template protection specifies \new{the properties (\new{irreversibility, unlinkability, and confidentiality})} which any template protection scheme must achieve to be deemed secure. \new{However, this evaluation methodology is not directly applicable to our work because anonymization seeks, on the contrary, \textit{to remove} the connection between identity and data.}

Le et al.~\cite{le_anonfaces_2020} discuss how to evaluate privacy-utility trade-offs for face anonymization, but their focus is \new{exclusively} on measuring the utility and not privacy.

Recent works~\cite{tekli_framework_2019, hao_robustness_2020, Fantomas} propose attacks on biometric data anonymization that use machine learning to reverse the obfuscation of images. These results show that the method is highly effective even when a human observer cannot recognize anything at all in the image. \new{The reversal of anonymization is indeed comparable to the training of a recognition system on anonymized data. However, we consider training recognition systems on anonymized data the more straightforward way to test whether identifying information remains in the anonymized data. Further, we also consider the reduction of the dataset.}

In the context of the {VoicePrivacy} challenge~\cite{tomashenko_introducing_2020}, other recent works have investigated the evaluation methodology of speaker anonymization. Noé et al.~\cite{noe_towards_2022} also propose a framework to evaluate and compare speech pseudonymization approaches using ZEBRA ~\cite{nautsch_privacy_2020} and voice similarity matrices~\cite{noe_speech_2020}. ZEBRA aims at creating a worst-case metric to evaluate speaker anonymization and voice similarity matrices allow to compare how well specific identities are anonymized. Bonastre et al.~\cite{bonastre_benchmarking_2021} propose a benchmarking methodology to test speaker recognition against spoofing and anonymization. \new{We investigate whether some of the methodological improvements to the evaluation of speaker anonymizations, like training recognition systems with anonymized data, can be applied to a wider range of biometrics like face and gait data.}


\new{In sum, many improvements to the evaluation methodologies of different research fields have been proposed. However, for the anonymization of biometric data, we find that multiple improvements can still be made to evaluation methodology, such as anonymized data in the training dataset and a more challenging anonymization scenario.}

\iffalse
Biometric recognition spans dozens of biometric traits and hundreds of techniques, but the methodology for evaluating the performance of these techniques has been assessed by only a very few works.

Goga et al.~\cite{goga_reliability_2015} assess the methodology for evaluating matching techniques of profiles from different social media platforms. They find that evaluation commonly overestimates the performance of the approaches by using an unrealistic methodology. Decann and Ross~\cite{DeCann_2013} investigate how the receiver operating characteristic (ROC) curve can be related to the \new{Cumulative Match Characteristic (CMC)} curve, finding that both ROC and CMC curves should be reported for biometric recognition. Granger and Gorodnischy~\cite{granger_evaluation_2014} describe the methodology that should be applied to evaluate the performance of biometric recognition for video surveillance applications. For the evaluation of stylometric authorship attribution, Stolerman et al.~\cite{stolerman_classify_nodate} make the case that an open-set model should be applied since in a realistic scenario the actual author might not be on the suspect list. Brennan et al.~\cite{brennan_adversarial_2012} propose adding attacks to the methodology of stylometry evaluation because most methods cannot defend against attacks.

More generally, Arp et al.~\cite{arp2022and} investigate the methodology of using machine learning in computer security. They describe common pitfalls of using the methodology for research purposes and also give advice on how to avoid the pitfalls.

Template protection is a very specific kind of privacy protection because it removes all possible attribute inferences while still allowing identity verification. One specialized evaluation methodology~\cite{rane_standardization_2014} for template protection specifies \dan{the properties (such as irreversibility and unlinkability)} which any template protection scheme must achieve to be deemed secure.

Le et al.~\cite{le_anonfaces_2020} discuss how to evaluate privacy-utility trade-offs for face anonymization, but their focus is \dan{decidedly different to ours because they focus exclusively }on measuring the utility.

Recent works~\cite{tekli_framework_2019, hao_robustness_2020, Fantomas} propose attacks on biometric data anonymization that use machine learning to reverse the obfuscation of images. These results show that the method is highly effective even when a human observer cannot recognize anything at all in the image.

In the context of the {VoicePrivacy} challenge~\cite{tomashenko_introducing_2020}, other recent works have investigated the evaluation methodology of speaker anonymization. The VoicePrivacy challenge, begun in 2020, aims at offering a common framework and methodology for evaluating speaker anonymization. Noé et al.~\cite{noe_towards_2022} also propose a framework to evaluate and compare speech pseudonymization approaches using ZEBRA ~\cite{nautsch_privacy_2020} and voice similarity matrices~\cite{noe_speech_2020}. ZEBRA aims at creating a worst-case metric to evaluate speaker anonymization and voice similarity matrices allow to compare how well specific identities are anonymized. Bonastre et al.~\cite{bonastre_benchmarking_2021} propose a benchmarking methodology to test speaker recognition against spoofing and anonymization.

\else
\fi



\section{Background}\label{sec:background}
In this Section, we define basic terminology required for our work.

\subsection{Biometrics, Inference, and Recognition}
\textit{Biometric traits} (also called biometric characteristics~\cite{ISO2382}) are properties of a human that \new{either capture the physiology} of a human (e.g. face, iris, fingerprint) or \new{its behavior} (e.g. voice, gait, heartbeat). \textit{Soft biometric traits} (e.g. age, sex, weight) are insufficiently entropic to positively identify an individual. However, the combination of soft biometrics can suffice to identify an individual.

Due to the unique nature of biometric traits for each human being they can be used for privacy-invasive inferences.
We distinguish between two privacy threats\new{. By the term \textit{identity inference} we mean that the identity of an individual is inferred. By the term \textit{attribute inference} we mean that only a specific private attribute (e.g. age, sex, medical condition) is inferred.}

In biometric recognition, identity \new{inference }and attribute inference are made operative in a system that learns an inference on representative samples for each class. For each biometric sample to be classified, a biometric recognition system returns a list of possible classes, where each class has been assigned its own separate likelihood. In closed-set recognition, the sample must belong to one of the classes in the dataset, while in open-set recognition the sample may belong to an unknown class.


\subsection{Anonymization} 
The aim of \textit{anonymization} is to protect an individual's identity. During the process of anonymization, information is removed or perturbed that is specific to an individual. Hence, anonymization prevents an adversary from using the data to infer the class corresponding to an individual (i.e. identification). In contrast to anonymization, \textit{pseudonymization} is aimed at retaining some connection between identity and data in order to link the data to an alternative identifier.
\new{Besides preventing identification, an anonymization or an pseudonymization also seeks to retain \textit{utility} (i.e. usefulness) of the data. Most often a trade-off between privacy and utility must be made.}

\newnew{An example scenario where biometric anonymization is used is the publication of images in newspapers where the identity of the person in the image should be protected. The basic ways to achieve anonymization here are to remove the identifying information (e.g., cropping the face out of the image), to coarsen the identifying information (e.g., pixelating the face), or to perturb the identifying information (e.g., adding noise to it). In most cases, it is not necessary to completely delete the identifying information, but rather to delete enough so that the person cannot be uniquely identified.}



\iffalse
\section{Survey of State-of-the-Art Evaluation Methodology for the Anonymization of Biometric Data}\label{sec:state-of-the-art}

In order to learn about the current state-of-the-art for evaluating biometric anonymization, we perform a survey study on 49 papers. We first explain our methodology for selecting and filtering the corpus of papers and then the taxonomy which we use to categorize them. After reporting the results of the survey we discuss the current state-of-the-art of evaluation methodology for biometric data anonymization. 

\subsection{Survey methodology}

For our survey, we reuse the corpus of papers \jt{of paper\textbf{s} ?}from two existing surveys on biometric data anonymization. The majority of papers are from Hanisch et. al.~\cite{hanisch_privacy-protecting_2021} \jt{does this reference need updating?} which collected papers that perform behavioral data anonymization and includes traits like voice, gait, and brain activity. Additionally, we use the corpus of face anonymization papers from a survey by Ribaric et. al.~\cite{ribaric_-identification_2016} \jt{if we were to do this now, a much more obvious choice would be Meden et al (2021) - obviously redoing this is out-of-scope now, but might be a critique point for reviewers} which focuses on anonymization in media content. From this initial corpus of papers, we then filtered the papers that do not fit our anonymization scenario (cf. Section \ref{sec:scenario&attacker}) in which the data is first anonymized and then published. This excludes methods like secure multi-party computation and offloading computing tasks via homomorphic encryption. Further, we also removed all papers that did not perform an empirical privacy evaluation of their method. This resulted in a final corpus of 49 papers, which are listed in Table~\ref{tab:list_survey_papers}. We then applied our taxonomy, which we describe next, to the corpus.

\subsection{Taxonomy}

The first category by which we separate the corpus is the \textit{biometric trait} the anonymization tries to protect and its \textit{protection goal}, which can be to prevent identity disclosure or attribute disclosure. \jt{are they actually calling methods preventing attribute disclosure "anonymization"?}
Since the anonymization approaches are tested against a biometric recognition system we note if the evaluations rely on a single approach or test \textit{multiple recognition systems}. 
Further, we examine if \textit{multiple parameters} for the anonymization technique are evaluated. Our main interest in this survey was to learn which kind of attacker model the evaluations employed. For this, we compare if an \textit{open-set} or \textit{closed-set} model was applied and with which kind of \textit{training data} (clear or anonymized) the recognition system was trained. Further, we check if the \textit{reversibility} of the anonymization approach was tested. As the last aspect, we compare the different \textit{metrics} employed to measure the anonymization performance.

\section{Results}

We find that \dan{most of these papers reviewed by us} focus on anonymizing voice data, then face, gait, and hand. 
Only two papers each tackle brain activity and eye-gaze (see Table~\ref{tab:trait}). Most papers try to protect against identity inference, six against attribute inference and five against both. 
Regarding metrics, we find that accuracy (we also include metrics that are closely based on accuracy e.g. $1 - accuracy$) is the most commonly used metric, followed by the EER, for measuring the anonymization performance. 
Some uncommon metrics we observed \dan{were} the usage of F1-Score~\cite{matovu_jekyll_2018}, and half total error rates (HTER)~\cite{matovu_your_2016} 
As seen in Table~\ref{tab:overview} a bit more than half of the papers evaluate different parameter configurations for their anonymization technique, while only about one in four papers uses more than one recognition system for its evaluation. For the set model we find that most papers use a closed-set approach. When it comes to training the recognition system all papers use clear data for training while only a minority also trains the recognition system with anonymized data. For the test, if the anonymization technique can be reversed we find only one paper~\cite{qian_hidebehind_2018} consider this for the evaluation, although it only performed a theoretical analysis.

\begin{table}
\centering
\begin{tabular}{c|p{5.5cm}}
Trait & Papers (Count and Sources)\\
\hline
Voice & 22 (\cite{abou-zleikha_discriminative_2015},~\cite{bahmaninezhad_convolutional_2018},~\cite{aloufi_emotionless_2019},~\cite{hamm_enhancing_2017},~
\cite{srivastava_evaluating_2020},~\cite{pribil_evaluation_2018},~\cite{nelus_gender_2018},~\cite{qian_hidebehind_2018},\newline~\cite{lopezotero_influence_2017},~\cite{parthasarathi_lp_2011},~\cite{keskin_measuring_2019},~\cite{pobar_online_2014},~\cite{nelus_privacy-aware_2019},~\cite{hashimoto_privacy-preserving_2016},~\cite{portelo_privacy-preserving_nodate},~\cite{fang_speaker_2019},\newline~\cite{justin_speaker_2015},~\cite{qian_speech_2021},~\cite{cohen-hadria_voice_2019},~\cite{jin_voice_2009},~\cite{vaidya_you_2019})\\

Face & 10 (\cite{neustaedter_blur_2006},~\cite{meng_face_2014},~\cite{gross_integrating_2005},~\cite{gross_model-based_2006}, ~\cite{newton_preserving_2005},~\cite{gross_face_2009},~\cite{meng_retaining_2014},\newline~\cite{boyle_effects_2000},~\cite{korshunov_using_2013},~\cite{korshunov_using_2013-1})\\

Gait & 8 (\cite{agrawal_person_2011},~\cite{hirose_anonymization_2019},~\cite{ivasic-kos_person_2014}, ~\cite{journew_toward_2018},~\cite{matovu_jekyll_2018},~\cite{tieu_rgb_2019},~\cite{tieu_approach_2017},~\cite{tieu_spatio-temporal_2019})\\

Brain Activity & 2 (\cite{yao_improved_2019},~\cite{matovu_your_2016})\\

Eye-gaze & 2 (\cite{steil_privacy-aware_2019},~\cite{bozkir_differential_2020})\\

Hand & 5 (\cite{leinonen_preventing_2017},~\cite{maiorana_bioconvolving_2011},~\cite{migdal_keystroke_2019},~\cite{monaco_obfuscating_2017},~\cite{vassallo_privacy-preserving_2017})\\
\end{tabular}
\caption{Publications included in the state-of-the-art survey with corresponding trait}\label{tab:list_survey_papers}
\end{table}

\begin{table}
\centering
\begin{tabular}{ l | c c c c c c}
Trait& Voice & Face & Gait & Hand & Brain & Eye \\ 
& 22 & 10 & 8 & 5 & 2 & 2 \\  
\hline
Protection Target & \multicolumn{2}{c}{Identity}  & \multicolumn{2}{c}{Attribute}  & \multicolumn{2}{c}{Both}  \\ 
 & \multicolumn{2}{c}{38} & \multicolumn{2}{c}{6} & \multicolumn{2}{c}{5} \\ 
\hline
Metric & \multicolumn{2}{c}{Accuracy}  & \multicolumn{2}{c}{EER}  & \multicolumn{2}{c}{Other}  \\ 
 & \multicolumn{2}{c}{36} & \multicolumn{2}{c}{10} & \multicolumn{2}{c}{3}\\
\end{tabular}
\caption{Publication count for biometric trait, protection goal, and metric to evaluate the technique}
\label{tab:trait}
\end{table}

\begin{table}
\centering
\begin{tabular}{ l | c c c c c c}
 & \multicolumn{3}{c|}{Yes} & \multicolumn{3}{|c}{No} \\ 
 \hline
anonymized training data & \multicolumn{3}{c|}{8} & \multicolumn{3}{|c}{41}\\  
test reversability &  \multicolumn{3}{c|}{1} &\multicolumn{3}{|c}{48}\\
closed-set assumption &  \multicolumn{3}{c|}{38} & \multicolumn{3}{|c}{11}\\
multiple parameters &  \multicolumn{3}{c|}{28} & \multicolumn{3}{|c}{21} \\
multiple recognition systems &  \multicolumn{3}{c|}{12} & \multicolumn{3}{|c}{37}\\
\end{tabular}
\caption{The number of papers for the remaining categories}
\label{tab:overview}
\end{table}

\subsection{Discussion}\label{sec:survey_discussion}
Since we took the papers for face anonymization from a different survey \dan{than} the other traits the number of papers per trait only gives an overview of the distribution of traits in our survey and cannot be generalized to the distribution over all existing publications. 
However, even though the number of face papers might be different in the wild, we believe that the distribution is close enough to allow conclusions about the overall evaluation methodology.

We find a large problem in today's evaluation methodology is the strength of the attacker model which is assumed for the evaluation. As we have seen most papers do not retrain their recognition system with the anonymized data. 
The results hence do not necessarily reflect how effective the attempted anonymization really is when facing specifically trained recognition systems.
Consider for example an anonymization which reorders each data dimension by one position. \jt{see intro}
Since the data structure has changed, the trained recognition model would most likely no longer work and therefore report a high anonymization performance.
Retraining the model, however, would reveal that the anonymization ineffective, because all features remain.
Another problem can be that the anonymization only removes the set of features that was used by the trained recognition system but another set of features, which also allow identification, remains untouched. Therefore, we believe that it is necessary to retrain the recognition systems with anonymized data to achieve reliable anonymization measurements.

As explained in Section~\ref{sec:background}, anonymization methods must be irreversible. However, only few papers actually evaluate this property either by a security analysis or by a practical attack. 
As the recent success of reversing anonymized face images~\cite{tekli_framework_2019-1, hao_robustness_2020, mcpherson_defeating_2016} shows, the threat of reversing anonymizations is real and even possible when a human observer cannot make any sense of the images anymore. \jt{reference fantomas} 
As such, not testing reversibility again assumes an unrealistically weak adversary and is a shortcoming of the current methodology.

Most of the papers in the survey evaluate with an underlying closed-set assumption.
We endorse this choice, since a closed-set assumption implicitly represents a stronger adversary. 
In a closed-world scenario, the adversary would have a fixed group from which the adversary can pick a result and can simply pick the identities/classes with the highest likelihood. Considering that adversaries might have additional information to identify a target we consider it important that the anonymization is evaluated in a closed-set scenario as the attacker might already have a fixed list of suspects that she wishes to identify.
A closed-set assumption is also tightly coupled with the metric that is being used to measure the anonymization performance.

More than two-thirds of the papers in our survey only use a single recognition system for their evaluation, most of time an approach which is the state-of-the-art at the time the paper is published. We challenge the underlying assumption that the approach which performs best on clear data will also perform best on anonymized data. As different recognition systems vary in how features are extracted from the biometric samples and how the classification is performed we think that testing with multiple approaches is necessary to rule out that the tested anonymization technique only works well against the specific recognition method.

A general observation in our survey was that the performance evaluation methodology often closely resembles the one used for evaluating authentication approaches. \jt{authentication or simply recognition?}
This may not demonstrate the efficacy of the anonymization well, as the evaluation goal in this case is to test the possibility of impersonation, rather than preventing identification.
A challenging problem instance for authentication can indeed be a simple example for anonymization, as it implies testing of identities that are difficult to distinguish to begin with. 
We argue that to create a good evaluation for anonymization the underlying recognition task should be chosen to be easy, as this increases the challenge for anonymization and translates to assuming a strong adversary. \jt{i think this subsection is already pretty good}

\else
\fi

\section{Improving the Evaluation Methodology}\label{sec:evaluation_methodology}

In this Section, we aim to achieve a reliable evaluation methodology for the anonymization of biometric data.
Our premise is that an evaluation methodology for anonymization techniques should be pessimistic and assume a strong adversary based on the worst-case performance of the anonymization technique. \new{To improve the evaluation methodology for the anonymization of biometric data,} we proceed in two steps.

First, we \newnew{present our adversary model and then} analyze the shortcomings of the state-of-the-art for evaluating biometric data anonymization. Overall, we find that \new{the evaluation of anonymization performance has been uncritically adopted from the evaluation of recognition systems.}


Second, we \new{make} three suggestions for improvement. We suggest (1) that recognition systems be trained with anonymized data, (2) that anonymization performance be tested against different recognition systems, and (3) that evaluation datasets consist of recognition problems more challenging to anonymization performance.

\subsection{Adversary Model}\label{sec:scenario&attacker}
\newnew{We investigate the efficacy of anonymizing biometric data, in other words, of preventing biometric recognition.
Hence, we consider a scenario in which a user provides his biometric data to a service provider to receive some utility. 
Examples of this kind of service are step counters based on gait data (e.g. for exercise/activity monitoring), a medical service that analyses the heart rate of users, a social media platform that is being used to publish images of the user, or a website that tracks the mouse movements of a user. 
Privacy-minded users will try to protect themselves or others from privacy inferences and therefore anonymize their data before sending it to the service.}

\newnew{We consider an adversary who gets full access to the data set submitted to the service, either because the attacker actually is the service provider or because the service provider leaks the data set. The adversary's goal is to perform privacy  inferences on the data set under attack. 
For this, the adversary has access to a training data set that contains labeled biometric data and can be used to train a biometric recognition system for this task. 
For the training data set, we consider that it can consist of both clear data or anonymized data.
We believe that this is a realistic assumption since the adversary will likely learn about the applied anonymization (similar to Kerckhoffs's principle) and can then apply the anonymization to the training set. Alternatively, the adversary might use scraped anonymized data, for example, from social media.}

\subsection{State-of-the-Art Evaluation Methods for the Anonymization of Biometric Data}
We began by gaining an overview of the problems \new{of the} state-of-the-art evaluation methods. To this end, we assessed the papers covered \new{in} two recent surveys~\cite{hanisch_privacy-protecting_2021, ribaric_-identification_2016} on the topic of biometric data anonymizations. Next, to gain a closer perspective on the field of face anonymization, we analyzed works \new{published from 2018}~\cite{rajabi_impracticality_2021, fan_image_2018, john_let_2020, wang_videodp_2019, hukkelas_deepprivacy_2019, maximov_ciagan_2020, shan_fawkes_2020}, \new{and one work from 2005}~\cite{newton_preserving_2005}. We included as many works as we could find which appeared at \textit{USENIX Security}, \textit{Privacy Enhancing Technologies Symposium} (PETs), and \textit{Data and Applications Security and Privacy}. \new{As a recent} work~\cite{Le_StyleID_2023} of 2023 testifies to the persistence of the said methodological flaws to this day. \new{An expanded survey of the current methodology can be found in the Appendix~\ref{sec:state-of-the-art}.}

\new{Our survey shows that the methods for evaluating techniques of biometric recognition or anonymization use the same recognition systems, the same datasets, and the same evaluation scenarios. This unquestioned reuse of the same attacker model, dataset, and scenario is highly problematic and will undermine the reliability of any evaluation of anonymization performance. Our reasoning is as follows. In biometric recognition, an evaluation method presents challenging scenarios to the recognition system. Identities are hard to distinguish from one another, the number of identities to be distinguished is high, the biometric samples are poor in quality, an open-set scenario is used, and imposters are introduced to mislead recognition systems. However, in biometric anonymization by contrast, these same conditions do not pose a challenge. In fact, for example the high number of identities makes anonymization much easier, because the more identities we have, the more likely it is that for each identity there is another similar identity in the dataset. This makes it harder to distinguish between identities, which makes anonymization easier. We conclude that anonymization performance will not be accurately evaluated by methods designed to evaluate the performance of recognition systems.}

Our analysis shows that the reusing of evaluation methods from recognition and anonymization causes three main problems.

The first problem we identified is that \new{reuse of the scenario for the evaluation of recognition makes for an unrealistically weak adversary model for the evaluation of anonymization}. \new{Since in most papers the recognition system is trained on clear data and not on anonymized data (e.g.~\cite{Le_StyleID_2023, hukkelas_deepprivacy_2019, rajabi_impracticality_2021}), obviously the implicit assumption being made is that the adversary is unaware of the anonymization in place.} However, an adversary which is aware of the anonymization can adapt to the anonymization and thus \new{will present} a greater threat. Consider, for example, an anonymization that performs a deterministic block permutation on a face image. The \new{modification of the} data would most likely cause the trained recognition model to break down, \new{and therefore report a high performance}. That report, however, will be based on flawed premises and is false.

The second problem we identified is that most evaluation methods assume that the recognition model which works best on clear data will also be the best model for recognizing people in anonymized data (e.g. ~\cite{tieu_rgb_2019, bahmaninezhad_convolutional_2018, fang_speaker_2019}). We challenge this assumption. Recognition models are developed on clear data\new{. N}o consideration is given to tampering with the data. Therefore, we doubt whether the recognition model which works best on the clear data is also the best for anonymized data.

The third problem we identified is that the same datasets are used to evaluate anonymization as are used to evaluate recognition (e.g. ~\cite{Le_StyleID_2023, maximov_ciagan_2020, rajabi_impracticality_2021})\new{. Consequently,} anonymization techniques are evaluated almost exclusively on large numbers of identities. We argue that it is more challenging for anonymization techniques when there are low numbers of identities in the dataset. Furthermore, a low number of identities is more realistic because \new{biometric data seldom exists alone and additional individuating information (e.g. device ids, soft biometrics, etc.) can be used to further reduce the number of identities in the group.}


\subsection{Our Improvements to State-of-the-Art Evaluation Methods}

We use closed-set recognition for our general scenario \new{to have a stronger attacker}. \new{Our adversary possesses a list of identities and consequently may simply} test samples against the list to select the most likely identity for a given sample.

\begin{figure}[!h]
\centering
      \includegraphics[width=0.4\textwidth]{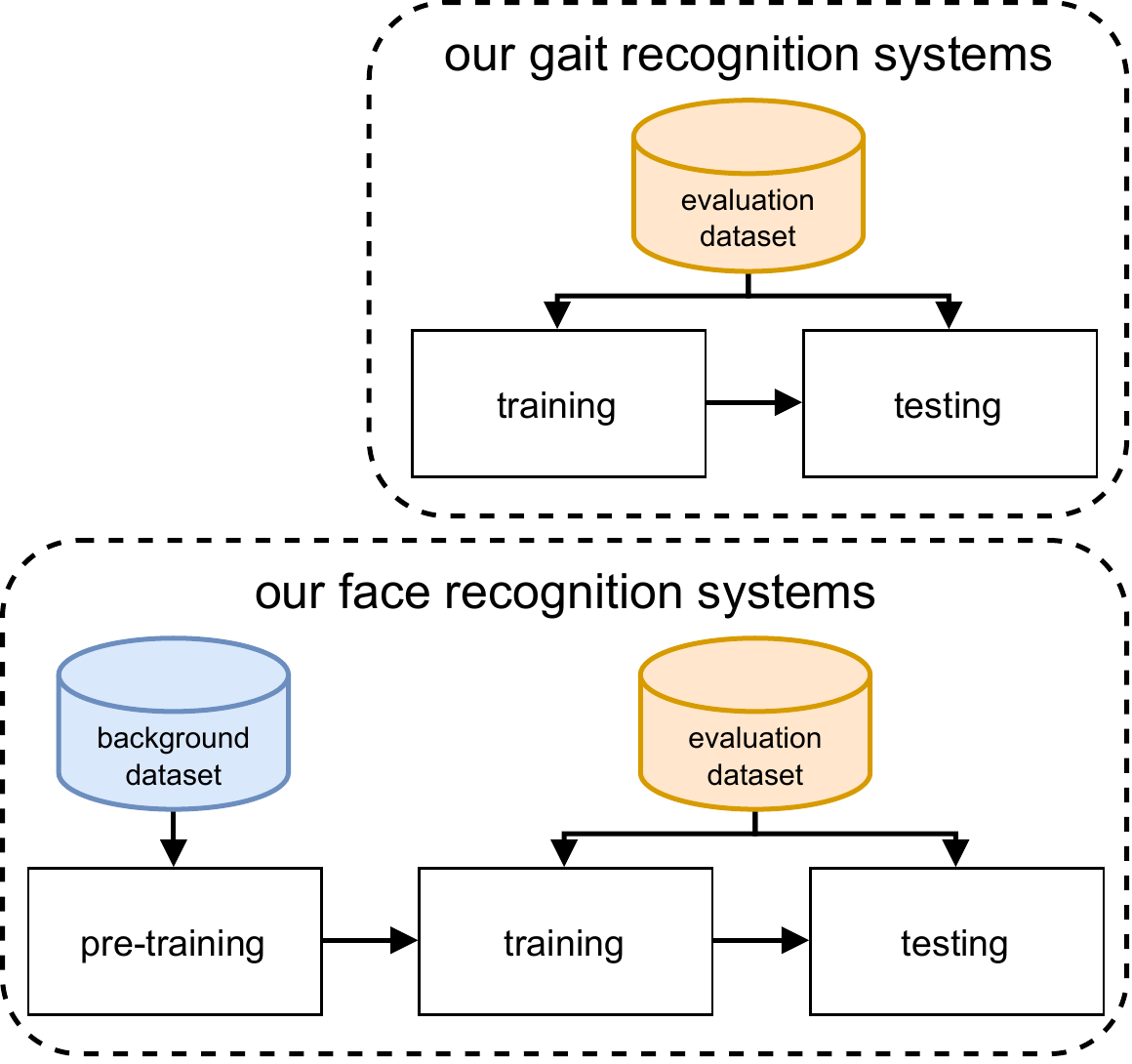}
\caption{Dataset use through the phases of our recognition systems for face and gait recognition.}\label{img:recognition-process}
\end{figure}

\new{We use two different biometric recognition system architectures for \new{the }gait and face recognition systems. For our gait recognition systems, we use an architecture which only uses data specific to the target identities, and for our face recognition systems, we use an architecture that uses additional background data not specific to the target identities (see Fig.~\ref{img:recognition-process}).
Both architectures split the samples of each identity contained in the evaluation dataset into \textit{train set} and \textit{test set}. The train set is used to learn a representation for each identity which is \new{then} used to infer the identity of the samples in the test set. \new{In addition} to this, the face recognition systems are \textit{pre-trained} \new{prior to} training on the train set. During pre-training, an additional background dataset \new{representative of} the general population is used to learn the features which can be used to differentiate between identities.}

\subsubsection{Training Recognition Systems with Anonymized Data}

In line with previous work~\cite{newton_preserving_nodate, mcpherson_defeating_2016, lal_srivastava_evaluating_2020, tomashenko_voiceprivacy_2021}, we propose that recognition systems be trained on anonymized data \new{so that a more reliable} anonymization performance \new{is achieved}. The idea of retraining recognition systems was first proposed for face recognition by Newton et. al.~\cite{newton_preserving_nodate}. Their model is trained with anonymized data and \new{then} tested on anonymized data\new{. The authors call this scenario} parrot recognition\new{, as opposed to training with clear data, which they call naive recognition}. The authors report much better performance for parrot recognition compared to naive recognition.

Parrot recognition is another term for \new{an} informed attacker, as defined by Srivastava et. al.~\cite{lal_srivastava_evaluating_2020}. In the evaluation of voice anonymization, Srivastava et. al.~\cite{lal_srivastava_evaluating_2020} propose three attackers who differ in their awareness of \new{the} anonymization. The ignorant attacker is unaware of the anonymization (as in black-box assumptions), the semi-informed attacker knows the anonymization algorithm (as in gray-box assumptions), and the informed attacker knows the algorithm plus the given parameters (as in white-box assumptions).

The VoicePrivacy challenge~\cite{tomashenko_voiceprivacy_2021, tomashenko_post-evaluation_nodate} used anonymized data to \new{train} a speaker verification system\new{. T}he system \new{was then tested }against anonymized voice samples. It was found that \new{training with} anonymized data already improved recognition performance; however, performance improvement was greater when the recognition was pre-trained with anonymized data. The results of the VoicePrivacy challenge show that (pre-)training the recognition system with anonymized data leads to a much stronger evaluation of the privacy performance of a technique. Therefore, we recommend training \new{and (where in use) also} pre-training recognition systems with anonymized data. But even when a complete pre-training of the model is not possible, \new{just} training with anonymized data can already \new{pose a more difficult challenge to an anonymization}.

\subsubsection{Test Against Different Recognition Systems}



Most evaluation methods rely on the state-of-the-art recognition system currently available for the targeted biometric trait. \new{However, during the design and development of recognition systems, anonymization is not considered. Consequently, recognition systems are not optimized to operate on anonymized data.
For this reason, we challenge the assumption that the state-of-the-art recognition systems will also be the one that performs best on the anonymized data. 
Obviously, for practical reasons, not all types of recognition systems can be used in an evaluation. However, at least a few conceptually different recognition systems should be tested in order to assess which techniques work best on the anonymized data. The aim here is to approximate worst-case performance of the anonymization.}

\subsubsection{Use a More Challenging Evaluation Dataset}\label{sec:security_margin}

\begin{figure}[!h]
\centering
      \includegraphics[width=0.475\textwidth]{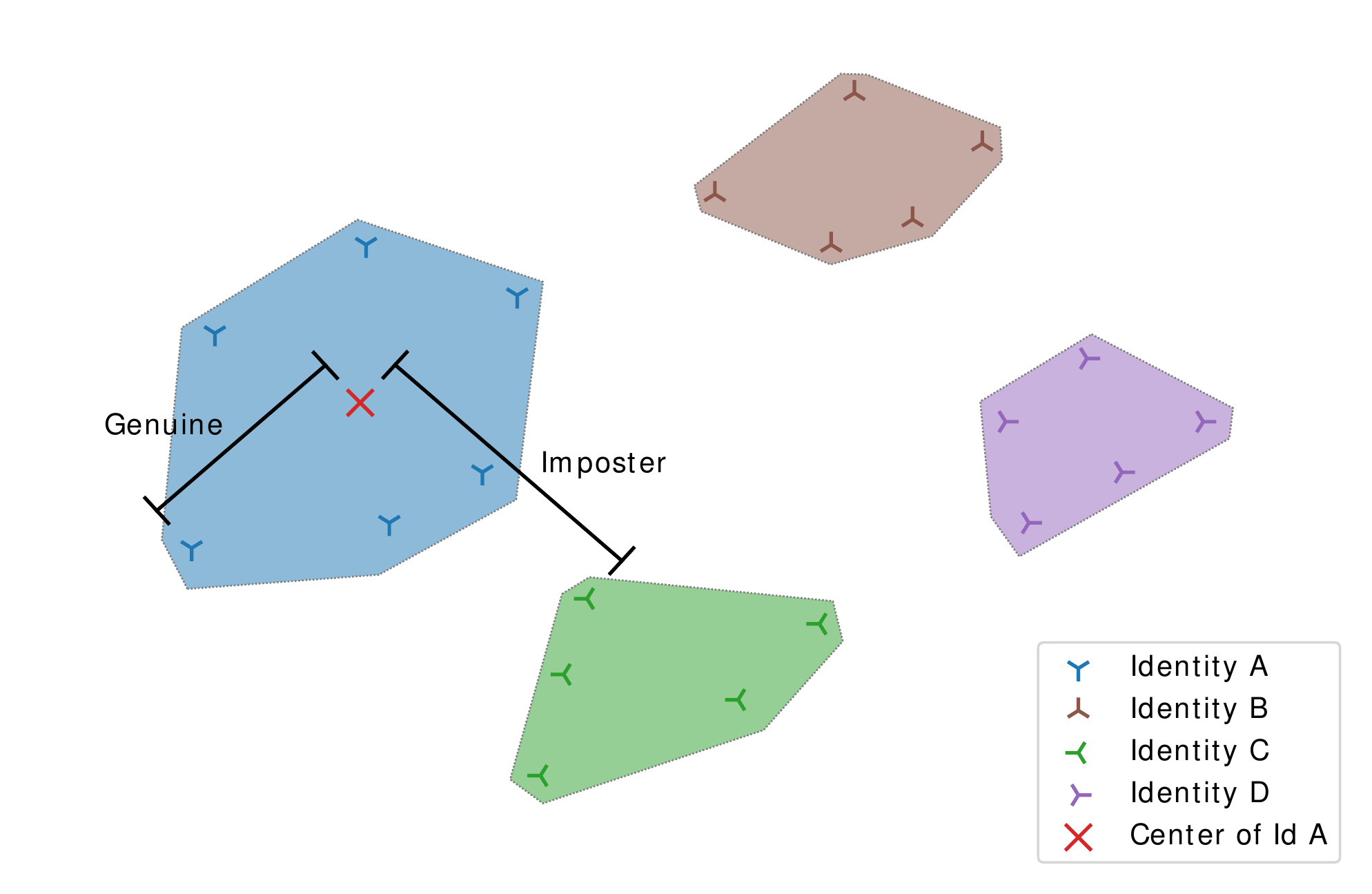}
\caption{Simplified example for Genuine and Imposter scores of an identity A in a 2D projection of the feature space.}\label{fig:distinctivemetrics}
\end{figure}

\new{T}he datasets currently being used for the evaluation of biometric recognition are\new{, as explained,} recorded and designed to pose challenging recognition problem. It is our proposition, though, that evaluators of anonymization use an easy recognition problem in order to create a challenging anonymization scenario. Since the recording of biometric datasets is time-consuming and expensive (not to mention complicated by \new{legal regulations like} GDPR), we propose that existing recognition datasets be \new{adapted} so that the easy recognition problem becomes a hard anonymization problem. In particular, instead of using the entire dataset, we propose that the identities in the dataset be reduced in number\new{. Further, we propose} that the selection of identities be based on the criterion of easy distinguishability. \new{For} the reduced dataset, our strategies for identity selection are as follows:

\begin{itemize}
    \item Random: As our baseline selection strategy, we use a random selection of identities\new{. We} repeat \new{the selection }multiple times to account for the variability of the selection.
    \item Classification: \new{We use a biometric recognition system on the anonymized data to select the identities which have the highest identification accuracy.}
    \item Metadata: We operationalize the fact that most biometric datasets also contain metadata about the identities, such as age and \new{sex}\new{. S}uch metadata \new{will typically }be \new{extractable} via a recognition system. \new{Our rationale is that identities with diverse attributes can be distinguished more easily when images are anonymized.} \new{Our procedure runs in three steps. }First, we normalize each point of metadata information between 0 and 1, and then we calculate the pair-wise Euclidean distance between the points. Second, we obtain a subset of identities by locating pairs of identities at the greatest distances from one another. And third, we calculate the average of distances between the identities in our subset, and then we consistently select the identity located at the maximum distance to the average.
    \item Feature-space: Many recognition systems work by projecting the biometric data into a feature space and then calculating distances between the feature vectors. The rationale is that the recognition system is trained to project datapoints from the same identity onto similar features and as well, to project datapoints from different identities onto contrasting features. However, misclassification occurs when the feature of a datapoint belonging to one identity is farther from the correct feature and closer to a feature belonging to another identity. Therefore, \new{we propose }that recognition performance be improved by the intentional selection of identities whose feature spaces are distant from one another \new{on anonymized data}. In other words, we choose identities who \new{are very different to} one another \new{when anonymized}. We use this idea to develop two selection strategies:
        \begin{itemize}
            \item Distinctive: Inspired by the Biometric Menagerie~\cite{Yager_Menagerie_2010}, we calculate for each identity a genuine score and an imposter score  (illustrated in Fig.~\ref{fig:distinctivemetrics}). The genuine score of an identity is the furthest Euclidean distance of any feature vector of this identity to the average of all feature vectors of this identity. The imposter score is the shortest Euclidean distance of the average of all feature vectors of this identity to a feature vector of any other identity. \new{Thus the \new{genuine} score is effectively an intra-class distance; conversely, the imposter score is effectively an inter-class distance. }If \new{the} inter-class distance is high \new{and the intra-class distance low}, then the identity is less likely to be misclassified because the features of other identities lie farther away. In sum, we select identities that have the best average of genuine and imposter scores.
            \item Center: Our purpose is to create a subset of identities lying at the greatest distances from one another. As with the metadata vector above, we begin by selecting the two identities whose average feature vectors have the largest Euclidean distance. Then we consistently select the identities whose average feature vectors lie at maximum distances from the average feature vector of our subset of identities.
        \end{itemize}
\end{itemize}

\section{Experiments}\label{sec:evaluation}

\new{Our evaluation is based on the physiological biometric face and behavioral biometric gait.} We begin by stating our hypotheses, and then we describe the experiments and present the results.


\subsection{Hypotheses}
Our aim in the evaluation is to test our three methodological proposals for improvements to the evaluation of biometric anonymization. \new{We have proposed, first, that recognition systems also be trained on anonymized data; second, that multiple recognition systems be used; and third, that a more challenging dataset be used.}

\new{We begin our testing by formulating five hypotheses:}
\begin{description}
\item[H1] \new{Training the recognition system on anonymized data achieves more reliable anonymization performance than training on clear data.}
\item[H2] \new{Training the recognition system on data in which a part of the samples is anonymized achieves more reliable anonymization performance than training on clear data.}
\item[H3] \new{No single recognition system simulates worst-case performance on all anonymizations.}
\item[H4] \new{A reduction in the number of identities in the evaluation dataset more robustly challenges the privacy protection of the anonymization.}
\item[H5] \new{The identities selected by our selection strategies are a more robust challenge to the privacy protection of anonymization.}
\end{description}

Our Hypotheses H1 and H2 hold that training recognition systems on anonymized data will achieve higher recognition performance. For our \textbf{H1}, we expect that (pre-)training recognition systems with anonymized data of the respective anonymization will result in higher recognition accuracies compared to (pre-)training on clear data. Further, for \textbf{H2}, we expect also that (pre-)training on partial anonymized datasets will perform better compared to (pre-)training on clear data. Further, we expect that increasing the amount of anonymized data in the train set will increase the recognition performance. We reason that the models we test must necessarily generalize more suitably to data that are noisier.


Our Hypothesis \textbf{H3} holds that no single recognition system will achieve the best performance on every anonymization. Our prediction for H3 is that, independent of results on clear data, some recognition systems will outperform others when using anonymized data. We reason that some recognition systems will better \new{learn features from the anonymized data.}


Our Hypothesis \textbf{H4} holds that reducing the number of identities in the evaluation dataset will present a more robust challenge to the performance of the anonymization. Our H5 builds on H4. For \textbf{H5}, we expect that selecting an evaluation dataset with our proposed selection strategies will pose a bigger challenge to the anonymization, and hence result in higher recognition performance then selecting random identities.






\subsection{Experiments}
We set an optimal performance bound by using chance-level performance of the anonymization as our baseline. We reason that perfect anonymization would leave adversaries with such a negligible advantage that their most effective strategy would be to guess identities at random. To approximate worst-case performance of the anonymization, we use the performance of clear level recognition, that is, the performance of the recognition system on clear data.

To test H1, we follow the same procedure for each anonymization technique: the recognition system is trained on the respective anonymized training data, and where \new{possible}, the system is also pre-trained on the anonymized data. To test H2, we (pre-)train the recognition system on different compositions of anonymized and clear training data using 25\%, 50\%, and 75\% anonymized training data. \new{Hence, we assess our H1 and H2 each with parrot and naive recognition.}

For our H3, we use different recognition systems and perform parrot recognition for each anonymization.




For our H4, we again perform parrot recognition. However, instead of using the full evaluation dataset, we use only a random subset of identities \new{of} 50\%, 25\%, 12.5\%, ..., until three of the original identities \new{remain}. For each number of identities, the sampling is repeated \new{ten} times to account for the variability of the random selection. Finally, in our last experiment for H5, we use the same numbers of identities as in the experiments for H4, but instead of randomly selecting, we choose identities \new{according to} the strategies described above in our methodology: Random, Classification, Metadata, Distinctive, and Center (see Subsection~\ref{sec:security_margin}). \new{We repeat the classification of the reduced dataset ten times to account for the randomness of the test/train split.}





\subsection{Datasets}
For the face recognition, we use the CelebA~\cite{liu_deep_2015} dataset because it is popular for face recognition and for anonymization evaluation\new{, and we use the WebFace260M~\cite{zhu2021webface260m} dataset because its images are realistic}. \new{From both datasets} we randomly select 1,000 identities \new{as evaluation set and another 9,000 identities as background dataset for retraining.} \new{We only select identities with at least eight images, and we limit the maximum number of images per identity to 20.} \new{We crop all images to the face region, with images containing multiple faces cropped to the largest face.}  We resize all images to 224x224 pixel \new{and rotate them until the eyes are level.}

\begin{figure}[!h]
    \centering
    \includegraphics[width=.3\linewidth]{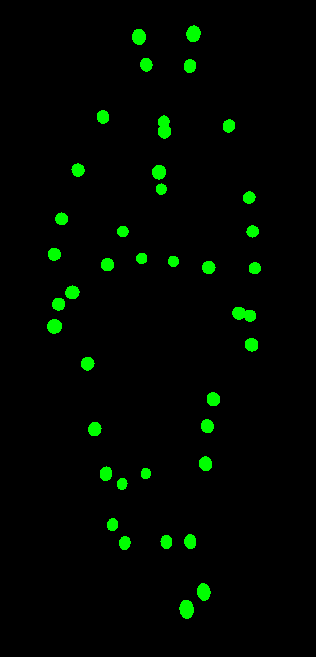}
    \caption{Sample pose of motion-captured gait information, represented as point-light walker.}\label{img:pointlight}
\end{figure}

For gait we use the dataset of the gait patterns of 57 identities by Horst et. al.~\cite{horst_explaining_2019}. The dataset represents the most comprehensive publicly available dataset that contains multiple gait samples per identity, and this, in particular, recommends the dataset to the evaluation of anonymization performance. For each identity in the dataset there are 20 gait sequences, and we resample these to be 100 frames long. The dataset has used optical markers to capture motion. The motion capture covers 52 tracked points, each given as absolute 3D position (see Fig.~\ref{img:pointlight}).

\subsection{Evaluation Framework}
\begin{figure}[!h]
\centering
       \includegraphics[width=0.45\textwidth]{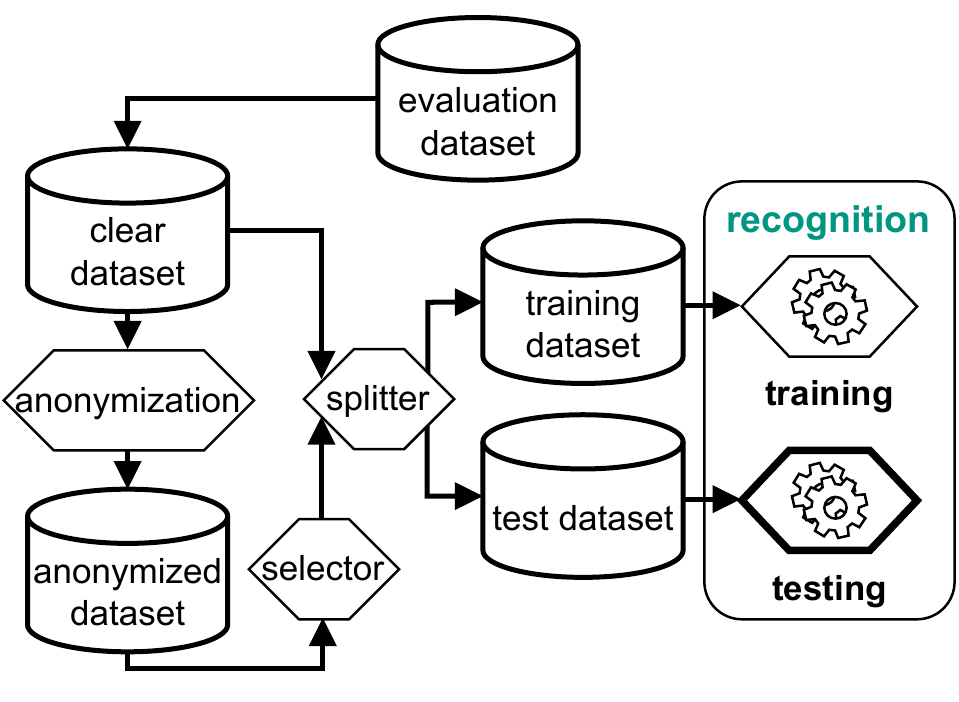}
\caption{Schematic overview of the evaluation framework architecture, excluding pre-training for simplicity}\label{img:framework_design}
\end{figure}

In order to run our experiments, we implemented the evaluation framework depicted in Fig.~\ref{img:framework_design}.

First, the clear dataset is copied and anonymized with a specific anonymization technique. Second, the selector performs a selection strategy to reduce the dataset to the configured numbers of identities. Third, the splitter splits the samples per identity into two sets, with 75\% of samples going into the train set and 25\% going into the test set. Depending on the configuration, either the clear samples or the anonymized samples go into the respective datasets.

Fourth and last, the \new{recognition system} is trained with the train set and evaluated with the test dataset. The resulting likelihood for a given test sample is recorded and saved for each identity.



\subsection{Recognition Systems}

For face recognition, we use the DeepFace~\cite{serengil_lightface_2020} library because it covers the entire face recognition \new{pipeline} and includes pre-trained models for ArcFace~\cite{deng_arcface_2019}, Facenet~\cite{schroff_facenet_2015}, and VGG-Face~\cite{parkhi_deep_2015}. Additionally, we use the face recognition model (frknn)~\cite{geitgey_face_2021}, which uses a pre-trained feature extractor \new{and k-nearest neighbors for classification}. \new{In order to also test non-deep-learning approaches, we use a scalar, principal component analysis (PCA) and support vector machines (SVM) pipeline as described in a scikit tutorial\footnote{\url{https://scikit-learn.org/stable/auto_examples/applications/plot_face_recognition.html}} and a recognition method that uses Google AI's mediapipe\footnote{\url{https://developers.google.com/mediapipe/solutions/vision/face_landmarker}} to extract 478 3-dimensional face landmarks before using a scalar, PCA and SVM pipeline on their coordinates.} \new{We also pre-train multiple models of ArcFace, which thereafter we referred to as Retrained ArcFace. For ArcFace pre-training, we used the remaining identities in CelebA \new{or WebFace260M} with the respective anonymization technique \new{under evaluation} applied to the samples. For \%-parrot recognition approaches, we anonymized only the \new{corresponding} percentage of the samples in the background dataset.
We validated Retrained ArcFace on clear data and achieved similar identification accuracy as the regular pre-trained ArcFace.}


For gait recognition, we use two types of feature vectors. The flatten feature vector simply flattens all poses of a gait sequence into a single vector, as proposed by Horst et al.~\cite{horst_explaining_2019}. The simple feature vector does a PCA over all poses of a walking sequence and then concatenates the 4 first components of the PCA with an average over all poses of the sequence. For classification, we use SVM, random forest, and k-nearest neighbors. Unless stated otherwise, we used the combination SVM+flatten for gait recognition.

\subsection{Anonymization Techniques}
In the following, we present the anonymization techniques we use for our evaluation. For face anonymization, we \new{select} simple anonymization techniques such as blurring and state-of-the-art machine learning anonymizations such as CIAGAN~\cite{maximov_ciagan_2020}. For gait anonymization, we use a subset of the anonymizations used by Hanisch et al.~\cite{hanisch2022understanding}. \new{If the anonymization is parameterized, we select the parameters in such a way that initially a low level of recognition accuracy is achieved. In this way, we can observe how our methodological improvements increase the recognition accuracy. Note that since we are investigating the efficiency of our methodological improvements, our selection of parameters does not allow a fair comparison of the anonymizations.}

\subsubsection{Face Anonymization}
\begin{figure}[!h]\centering\setlength{\tabcolsep}{1.5pt}
\begin{tabular}{cccc}
      \includegraphics[width=2cm]{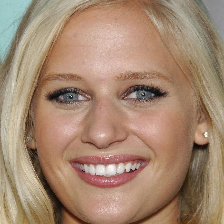} & \includegraphics[width=2cm]{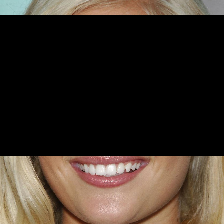} & \includegraphics[width=2cm]{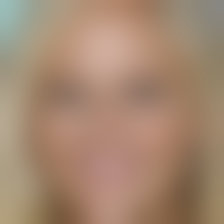} & \includegraphics[width=2cm]{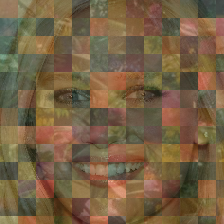}\\
      Original image & Eye Masking & Gaussian Blur & k-RTIO\\
      \includegraphics[width=2cm]{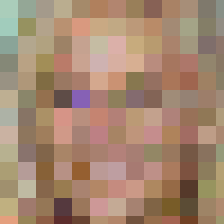} & \includegraphics[width=2cm]{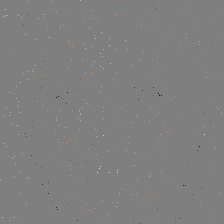} & \includegraphics[width=2cm]{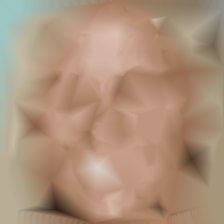} & \includegraphics[width=2cm]{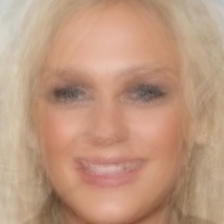}\\
       DP Pix & DP Snow & DP Samp & \textit{k}-Same-Pixel\\
      \includegraphics[width=2cm]{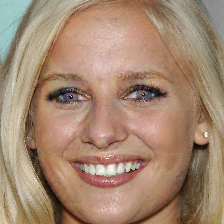} & \includegraphics[width=2cm]{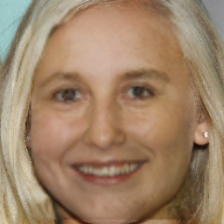} & \includegraphics[width=2cm]{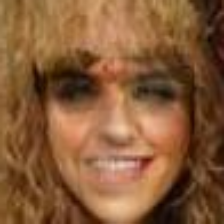}\\
      Fawkes & DeepPrivacy & CIAGAN
    \end{tabular}
\caption{Example image for each of the face anonymization techniques we assess.}\label{img:faceanon}
\end{figure}

We consider the following techniques for face anonymization in our evaluation (see Fig.~\ref{img:faceanon}). The \textit{Eye Masking} anonymization uses a black strip with 140 pixels height to cover the eye area of the face. \textit{Gaussian Blur} applies a gaussian blur with a kernel size of 101. The anonymization k-randomized transparent overlays (\textit{k-RTIO}) ($\alpha=0.4$, $blocksize=18$, $k=3$) by Rajabi et al.\cite{rajabi_impracticality_2021} add a block-permuted semi-transparent overlay to the face image. The three methods \textit{DP Pix}~\cite{fan_image_2018} ($\epsilon=2$, $b=12$, $m=16$), \textit{DP Snow}~\cite{john_let_2020} ($d=0.01$), and \textit{DP Samp}~\cite{wang_videodp_2019} ($\epsilon=5$, $k=24$, $m=12$) use differential privacy (DP) to provide formal privacy guarantees. We adapted these three methods from Reilly et al.~\cite{reilly_comparative_2021} for RGB images. \new{Our} adaptation to RGB images \new{prevents us from providing} the formal guarantees given for grayscale images. Another formal privacy framework is \textit{k}-anonymity, as used in the anonymization \textit{k-Same-Pixel} ($k=10$) by Newton et al.~\cite{newton_preserving_2005}. \textit{k}-Same-Pixel expects a static dataset with a single image per identity. This does not apply to our \new{scenario} because we anonymize image by image and have multiple images per identity. Therefore, we use a separate background dataset with 200 identities. This means that the formal guarantees do not apply to our implementation.
In Fawkes~\cite{shan_fawkes_2020} ($mode=high$), adversarial machine learning is used to poison face recognition training data and thereby protect \new{the identity in the picture}. Both \textit{DeepPrivacy}~\cite{hukkelas_deepprivacy_2019} and \textit{CIAGAN}~\cite{maximov_ciagan_2020} anonymize faces by replacing them with new synthetic ones and then fitting them into the original background.

\subsubsection{Gait Anonymization}
For \new{our gait experiments}, we use simple anonymization techniques to select precisely \new{the} information to \new{be perturbed }in the samples. First, we suppress parts of the samples: \textit{Keep(legs)} and \textit{Keep(head)} both keep only the captured points for legs or head, respectively, while all other points are set to zero. Second, we perturb the samples: \textit{Noise(x)} applies to each captured point normal ($\mu=0$, $\sigma=1$) distributed noise, which is scaled by 3, 10, or 100. Third, we generalize: \textit{Motion Extraction} captures the differences between each next pose in order to extract only the dynamic parts of the data. The structure of the walkers is, then, effectively removed.

\subsection{Selection Strategies}

\new{For our selections of face data using the Classification strategy, we use ArcFace to calculate the identification accuracy for each identity. We also use ArcFace to extract the feature vectors for the Center and Distinctive strategies. For gait, we use SVM+flatten for the Classification strategy and a PCA with four components over all samples as feature vector for Center and Distinctive.}

\subsection{Framework Implementation}

Our evaluation framework was implemented using python (version 3.8) with numpy (1.19.5), scikit-learn (0.23), and DeepFace\cite{serengil_lightface_2020} (0.0.65) libraries. 

\section{Results}\label{sec:results}

We report here the results of our experiments. We assess, in turn, the validity of each of our hypotheses: whether recognition systems trained on anonymized data improve evaluation performance (H1, H2), whether no single recognition system performs best on every anonymization (H3), and lastly whether a reduction in the number of identities \new{(H4)} and \new{whether a }selection of identities in the evaluation dataset \new{actually pose} real challenges to the \new{privacy protection} of the anonymization (H5). \new{In the Appendix~\ref{sec:additional_results}, there are additional results using the WebFace260M dataset and the  selection on clear data.}

\subsection{Recognition Systems Trained on Anonymized Data Improve Evaluation Performance}
\begin{figure}[!ht]
\centering
      \includegraphics[width=0.47\textwidth]{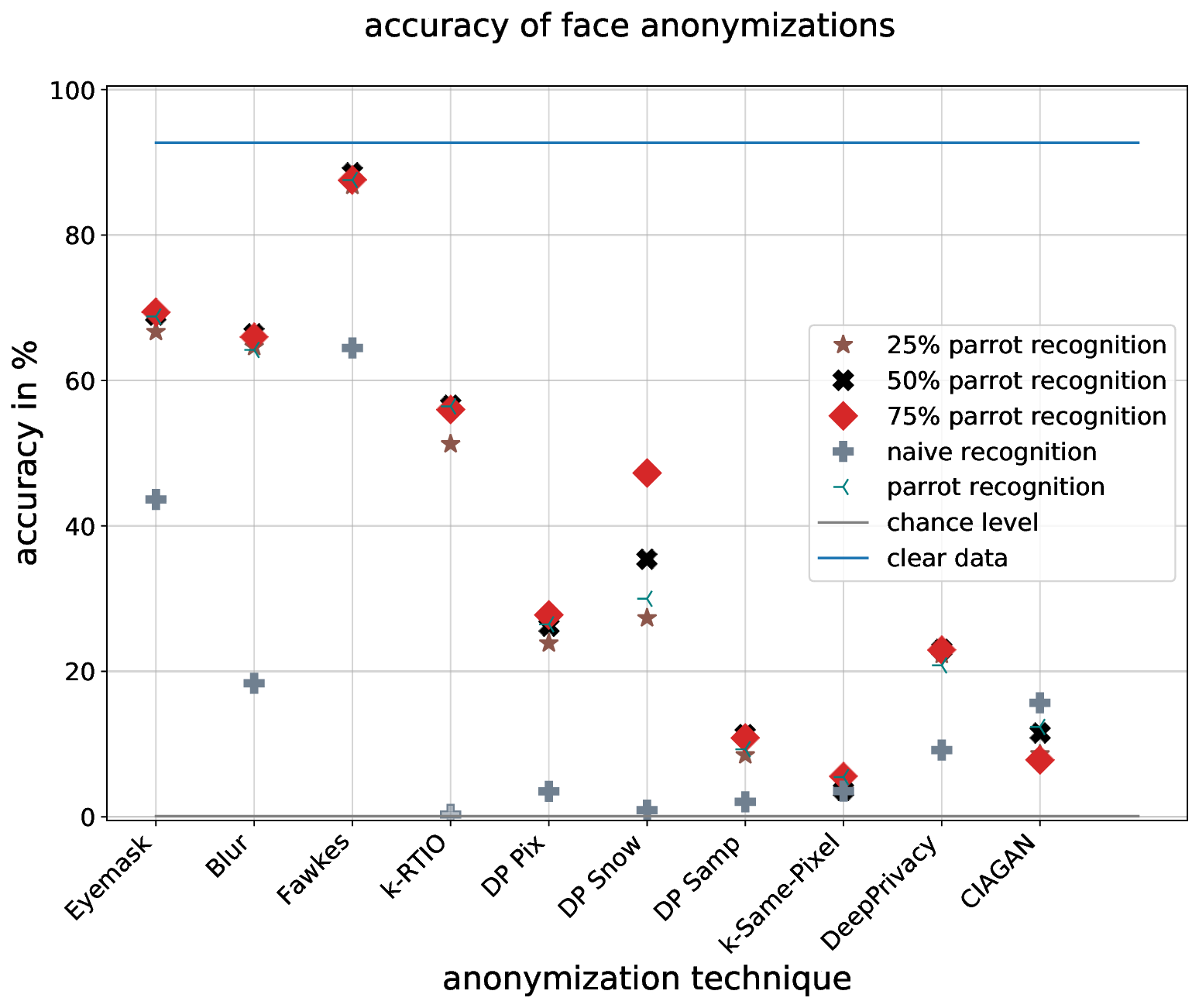}
\caption{Accuracy for face anonymizations using \new{ArcFace retrained} \newnew{on the CelebA dataset} with naive, \%-parrot, parrot recognition. A lower accuracy means better privacy protection.}\label{img:exp1_face}
\end{figure}

\begin{figure}[!ht]
\centering
      \includegraphics[width=0.47\textwidth]{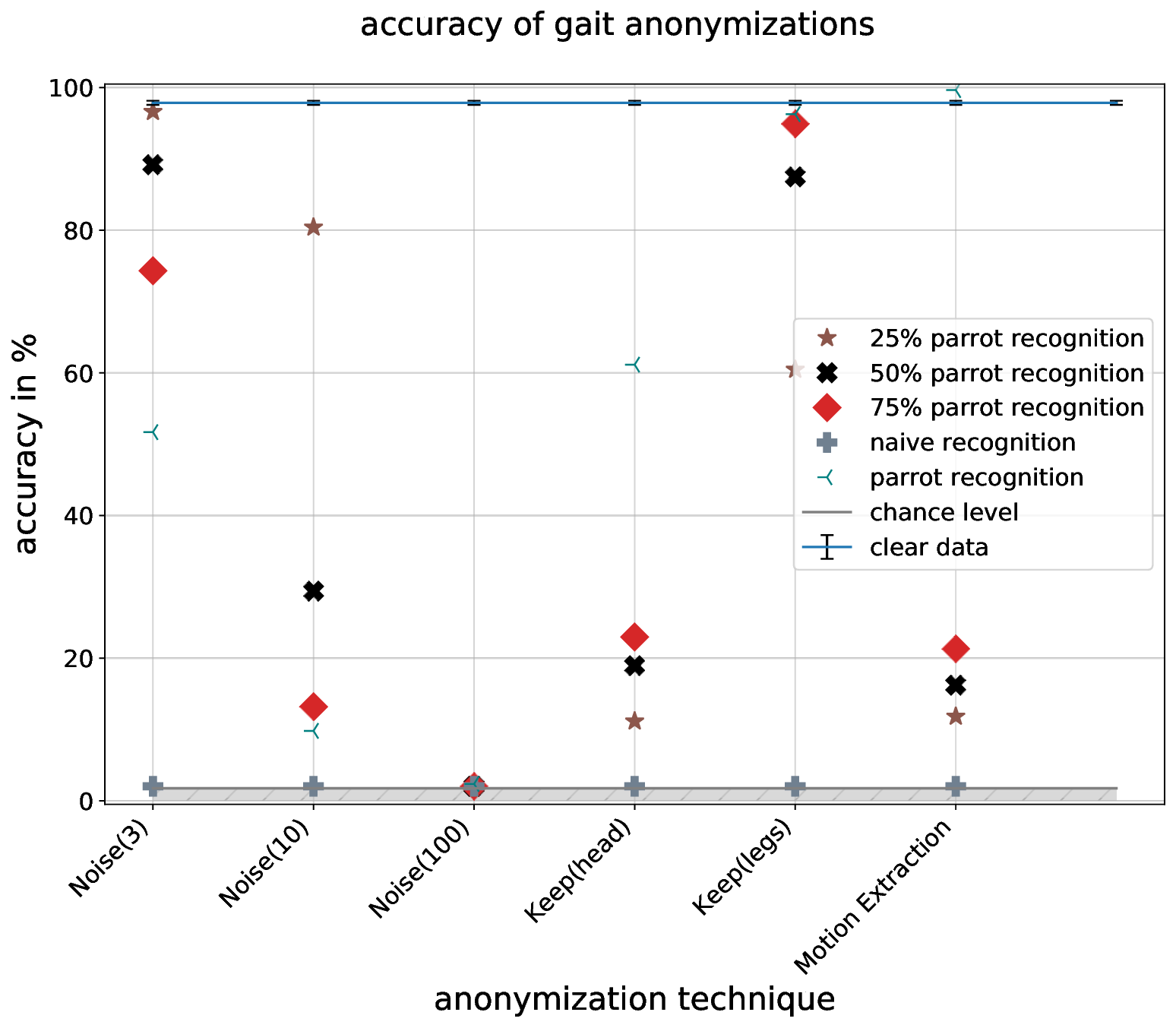}
\caption{Accuracy for gait anonymizations using SVM+simple with naive, \%-parrot, parrot recognition. A lower accuracy means better privacy protection.}\label{img:exp1_gait}
\end{figure}

In Fig.~\ref{img:exp1_face} and Fig.~\ref{img:exp1_gait}, we present the results of our experiments for H1 and H2 on the anonymization of face data and \new{for }gait data.

For face images, we find that, except for CIAGAN and \textit{k}-Same-Pixel, all parrot recognition systems perform better than naive recognition. \new{For \textit{k}-Same-Pixel, all recognition systems have nearly the same performance, while for CIAGAN, naive recognition performs best.} This anomaly \new{in CIAGAN }makes sense when we consider how CIAGAN performs the anonymization: every face is replaced by another face which shares the same soft biometrics. Therefore, we assume that CIAGAN's replacement of the face on each training image makes it harder \new{for} \new{ArcFace Retrained} to learn useful feature vectors.

We find significant results for parrot recognition of face anonymization. The performance of full parrot recognition and \new{of }all \%-parrot recognition cluster close together for most anonymizations. In fact, \%-parrot recognition often achieves the same performance as the full parrot recognition, and for DP Snow, the 75\%-parrot recognition even outperforms the full parrot recognition.

In contrast to our results for face anonymization, the results for gait anonymization show full parrot recognition outperforming \%-parrot recognition, with the exception of all Noise anonymization (cf. Fig.~\ref{img:exp1_gait}). For all \new{gait }anonymizations, naive recognition performs only at the chance-level. The \%-parrot  results for Noise(3) and Noise(10) are interesting because 25\% performs best, 50\% performs second best, 75\% performs third best, and full parrot performs worst.

In our results for both face and gait anonymization, one thing defied our predictions. In the face and gait anonymization of DP Snow, Noise(3), and Noise(10) anonymization performance improves when the model is trained solely on a portion of anonymized images rather than on the full anonymized training set. \new{We draw attention to the fact that all three anonymizations perform noise injection either by adding noise to each datapoint or by randomly removing pixels from the image. That portion of noisy data samples in the training set enables the recognition systems to adapt to DP Snow, Noise(3), and Noise(10) while still learning the features required for the classification from the clear data. We conclude, therefore, that there is a tipping point where more noisy data no longer improves training performance but begins impairing it.}

\subsection{No Single Recognition System Performs Best on All Anonymizations}
\begin{figure}[!ht]
\centering
      \includegraphics[width=0.47\textwidth]{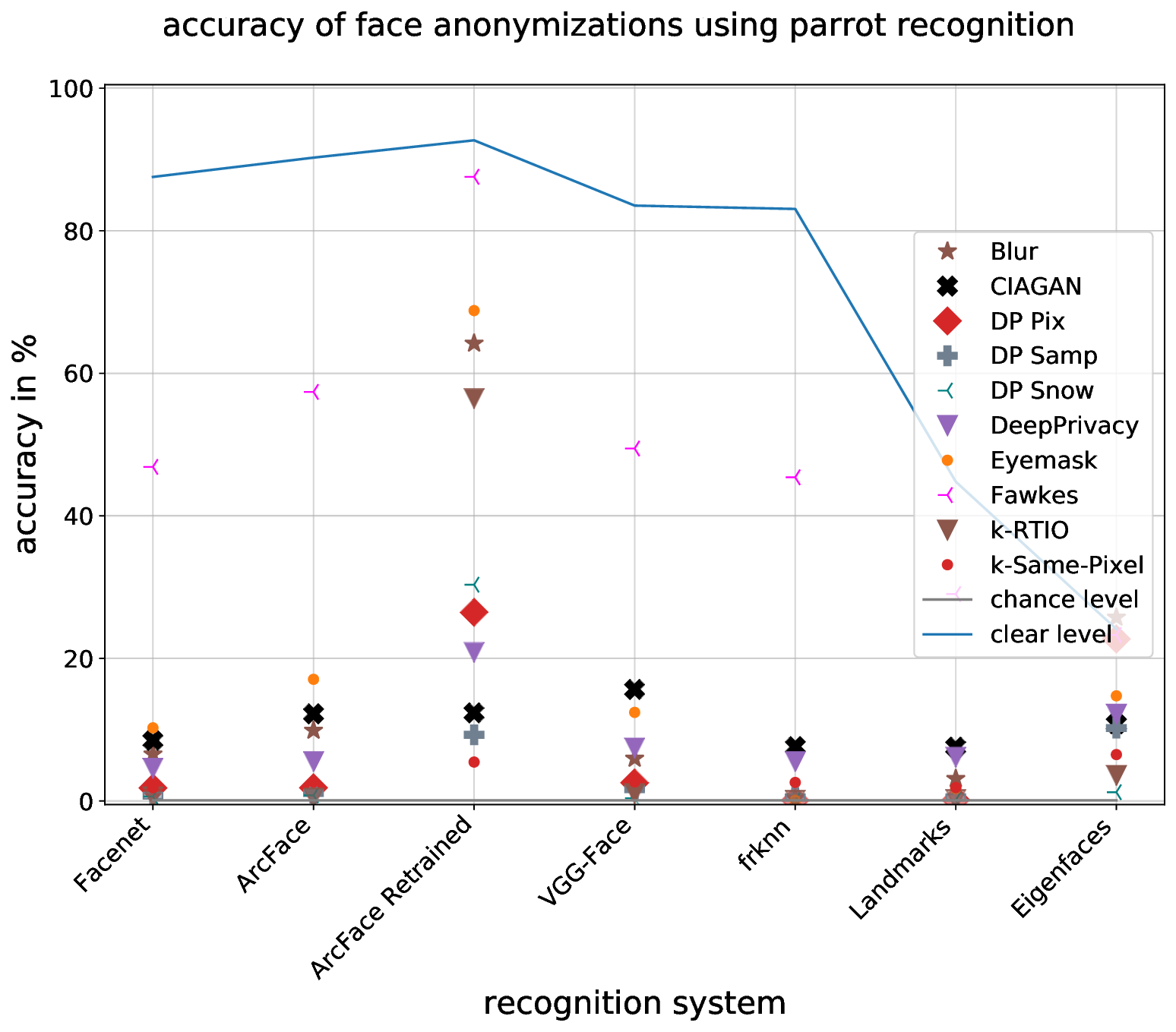}
\caption{Accuracy of face anonymization over different recognition systems using parrot recognition \newnew{on the CelebA dataset}. A lower accuracy means better privacy protection.}\label{img:exp3_parrot_face}
\end{figure}

\begin{figure}[!ht]
\centering
      \includegraphics[width=0.47\textwidth]{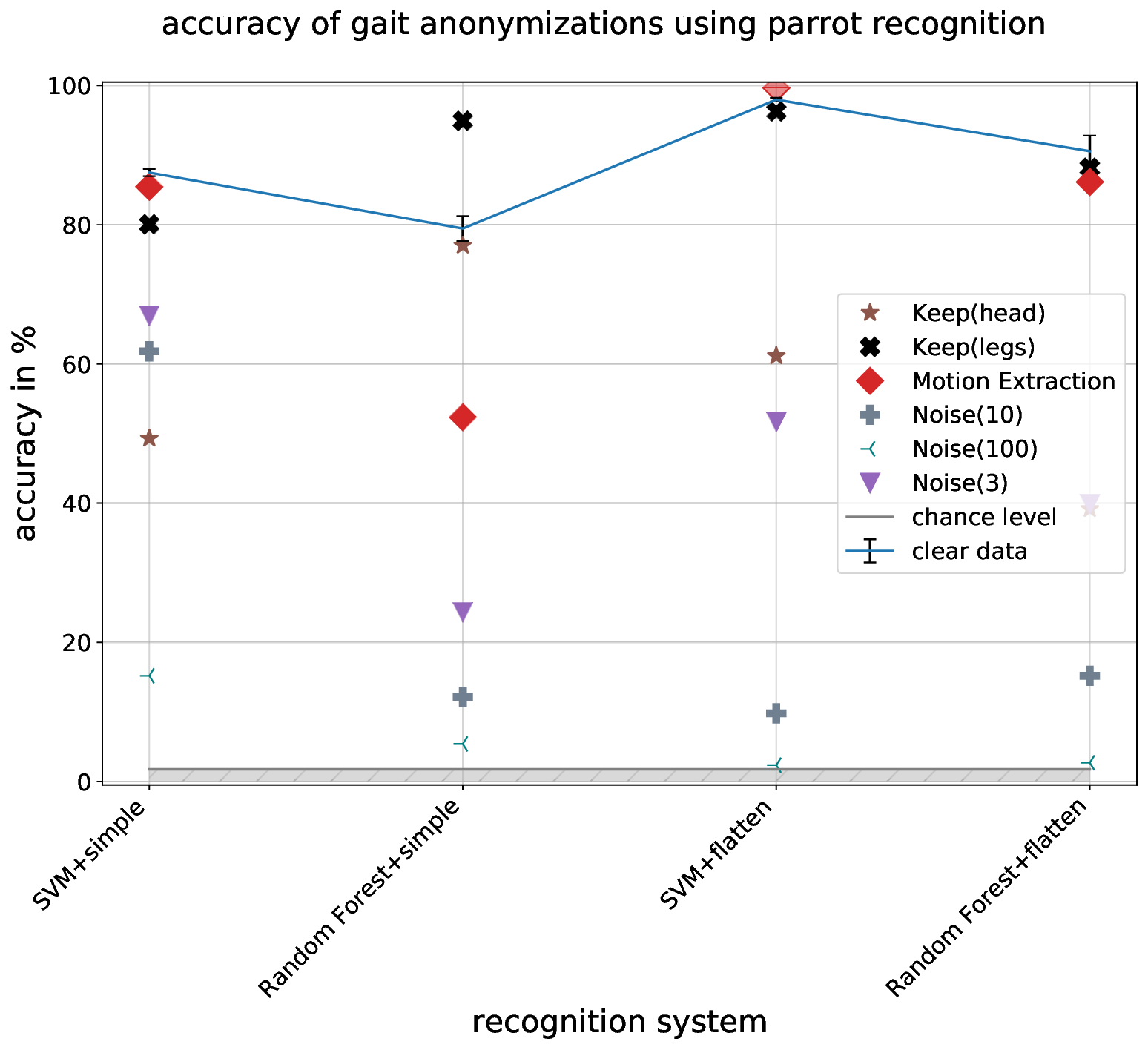}
\caption{Accuracy of gait anonymization over different recognition systems using parrot recognition. A lower accuracy means better privacy protection.}\label{img:exp3_parrot_gait}
\end{figure}

We present the results of our experiments for H3 for the anonymization of face data in Fig.~\ref{img:exp3_parrot_face} and for the anonymization of gait data in Fig.~\ref{img:exp3_parrot_gait}.

\new{All face anonymizations, except Fawkes, achieve a performance below 30\% for all recognition systems except ArcFace Retrained.} Fawkes achieves between \new{30\%} and 60\% \new{(except with Eigenfaces)}. The results for ArcFace Retrained differ significantly. With ArcFace Retrained, most anonymization techniques achieve much higher recognition rates\new{. O}nly CIAGAN, DP Samp, and \textit{k}-Same-Pixel are still below 30\%\new{, while} Blur, DP Snow, and Fawkes are even above 60\%. \new{An interesting observation is that while Eigenfaces performs worst on clear data it performs better on DP Pix and Blur than most other recognition systems.}

For the gait data, all combinations of techniques perform between 80\% and 98\% on clear data, with SVM+flatten performing best on the clear data. \new{The gait anonymization techniques across recognition systems perform in the same order, that is, we find the worst performance for Noise(100) and we find the best performance for either Keep(legs) or Motion Extraction.}

\new{We note} that the differences between the gait anonymization techniques across the recognition systems can be quite large. For example, SVM+simple Noise(100), Noise(10), and Noise(3) score much higher when compared to the \new{other} recognition systems. However, among the anonymizations that do not use noise injection, SVM+simple scores lower than SVM+flatten. In sum, we observe that no single gait recognition system outperforms the others.

\subsection{Reducing the Number of Identities in the Evaluation Dataset Increases the Challenge for the Anonymization}
\begin{figure}[!ht]
\centering
      \includegraphics[width=0.47\textwidth]{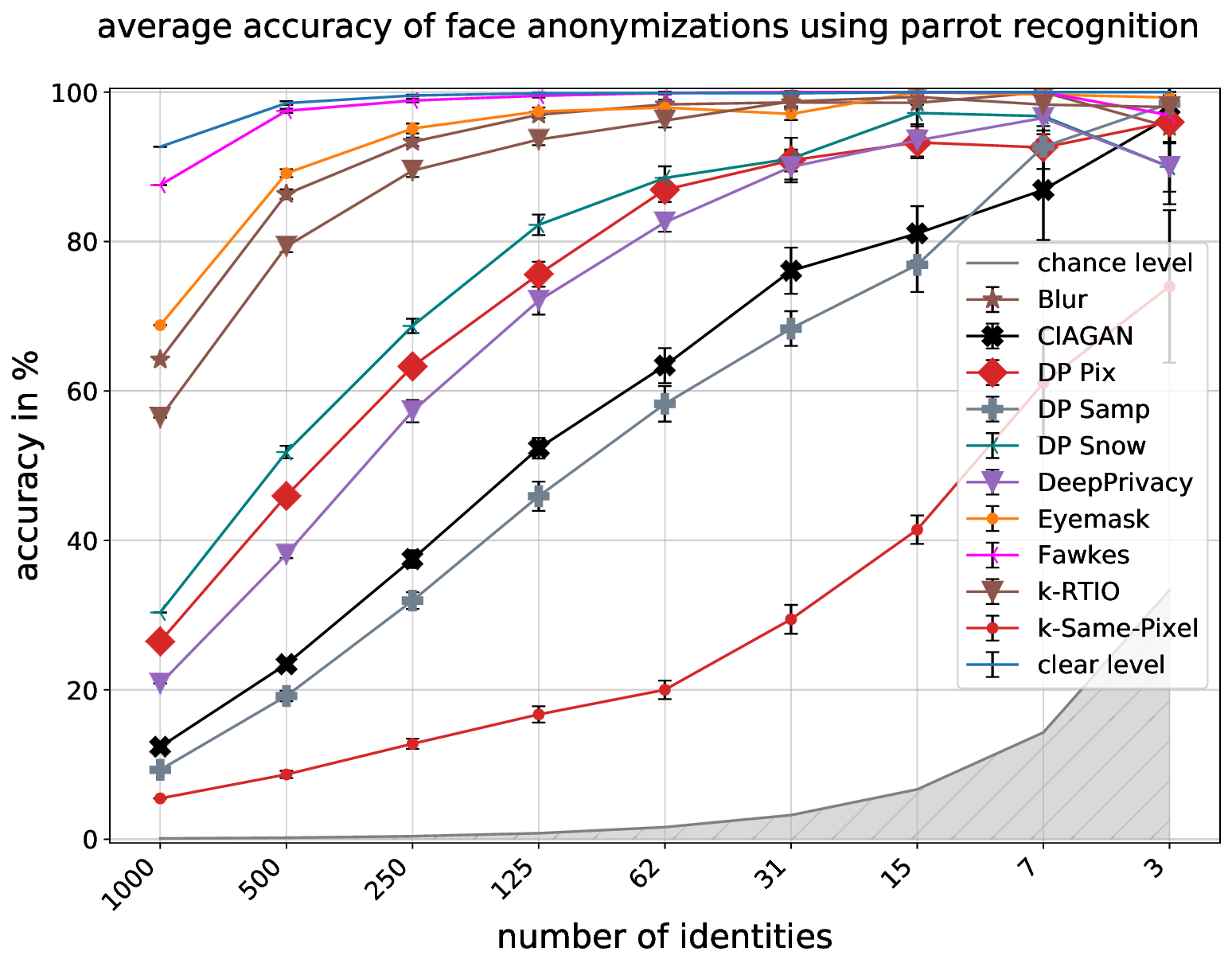}
\caption{Mean accuracy of face recognition over ten random selections \new{(excluding 1000 identities)} from decreasing numbers of identities. The \new{standard deviation} of the random selection is given as error bars. ArcFace Retrained is used with parrot recognition \newnew{on the CelebA dataset}. A lower accuracy means better privacy protection.}\label{img:exp2_face}
\end{figure}

\begin{figure}[!ht]
\centering
      \includegraphics[width=0.47\textwidth]{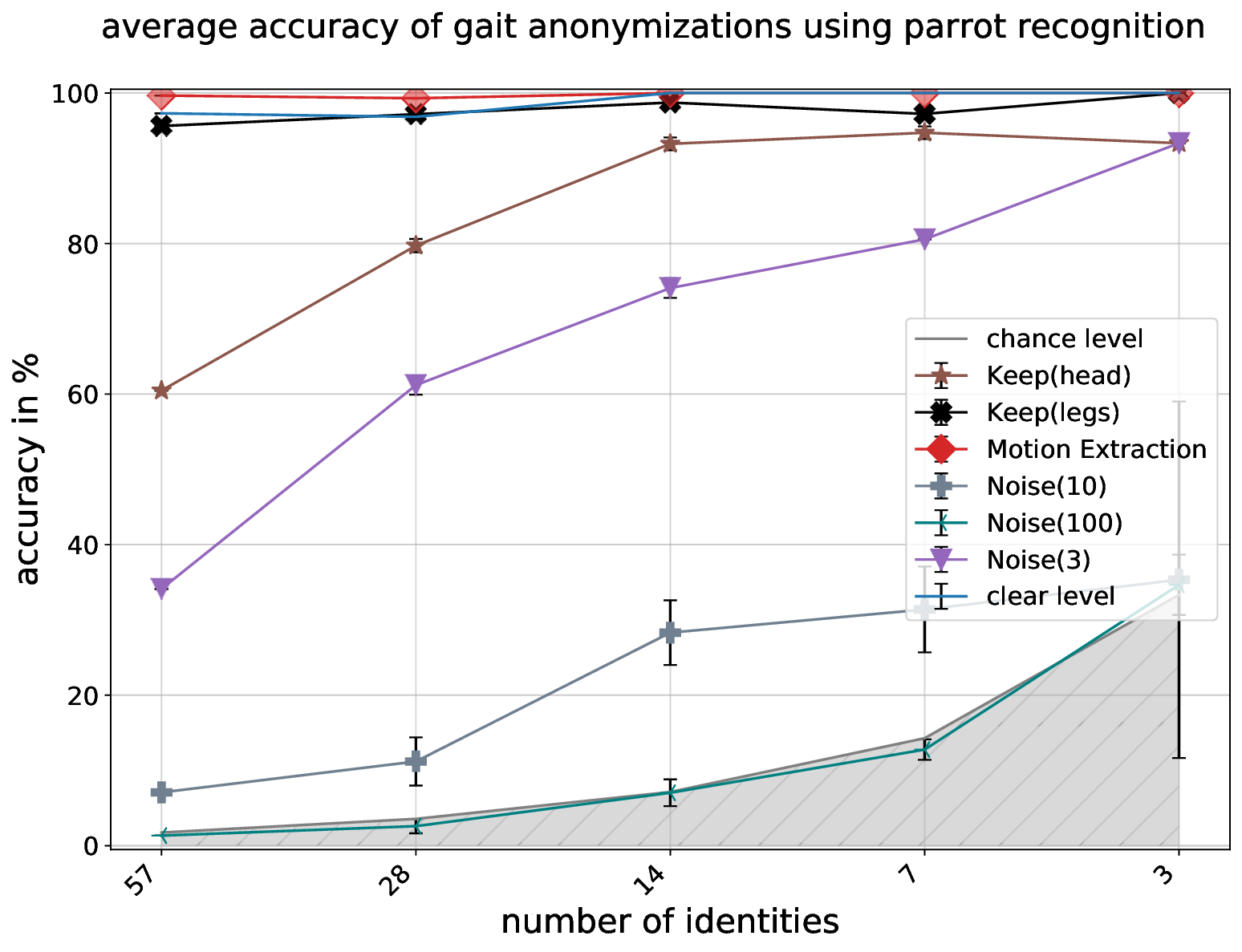}
\caption{Mean accuracy of gait recognition over ten random selections \new{(excluding 57 identities)} \new{for} decreasing numbers of identities. The standard deviation of the random selection is given as error bars. SVM+flatten is used with parrot recognition. A lower accuracy means better privacy protection.}\label{img:exp2_gait}
\end{figure}

 We present the results of our experiments for H4 for the anonymization of face data in Fig.~\ref{img:exp2_face} and for the anonymization of gait data in Fig.~\ref{img:exp2_gait}.
 
For the face data, we assess the accuracy of our H4 by comparing the performances of parrot recognition on different numbers of identities in the evaluation dataset (see Fig.~\ref{img:exp2_face}). For each number of identities (except \new{the number of }the full dataset), we selected 10 random subsets and calculated average performance and standard deviation. Every decrease in the number of identities increases the chance-level performance for the recognition systems. In short, the decreases make it easier for the recognition system to randomly guess an identity. \new{We observe this increase in performance for all anonymization techniques}. In particular, Fawkes attains the same performance plateau as initially on the clear data. Eyemask, Blur, and k-RTIO also start at high performance, but need longer to approach clear-level performance. \textit{k}-Same-Pixel is the best performing anonymization. \textit{k}-Same-Pixel stays close to the chance-level  while mimicking the same increase in accuracy. In sum, we observe that decreases in numbers of identities increase the \new{standard deviation} of accuracy\new{. From this, we reason that }the selection of identities for the evaluation group is an decisive factor in evaluation accuracy.

For the gait data (Fig.~\ref{img:exp2_gait}), we observe a similar increase in recognition performance, except for the anonymization techniques Noise(10) and Noise(100), which stay close to the chance-level . The techniques Noise(10) and Noise(100) increase the \new{standard deviation of the performance} as the number of identities decreases. \new{For the other gait anonymizations, we} do not observe \new{the same} relation in the \new{standard deviation}.

\begin{figure}[!ht]
\centering
      \includegraphics[width=0.47\textwidth]{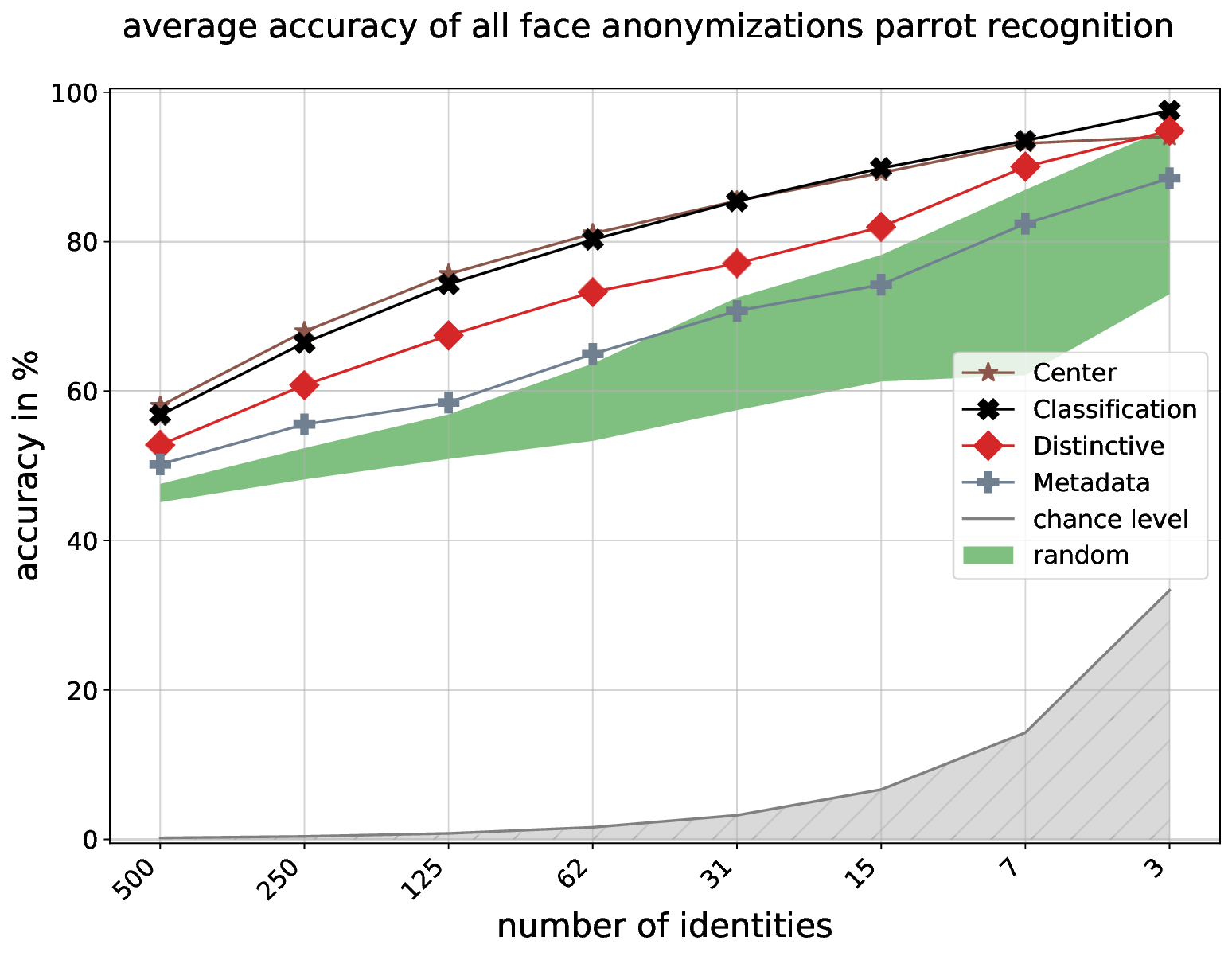}
\caption{All accuracies are given as the average across all face anonymization techniques. The green area indicates the accuracy range of the previous ten random selections of identities. ArcFace Retrained is used with parrot recognition \newnew{on the CelebA dataset}. A lower accuracy means better privacy protection.}\label{img:exp4_face}
\end{figure}

\begin{figure}[!ht]
\centering
      \includegraphics[width=0.47\textwidth]{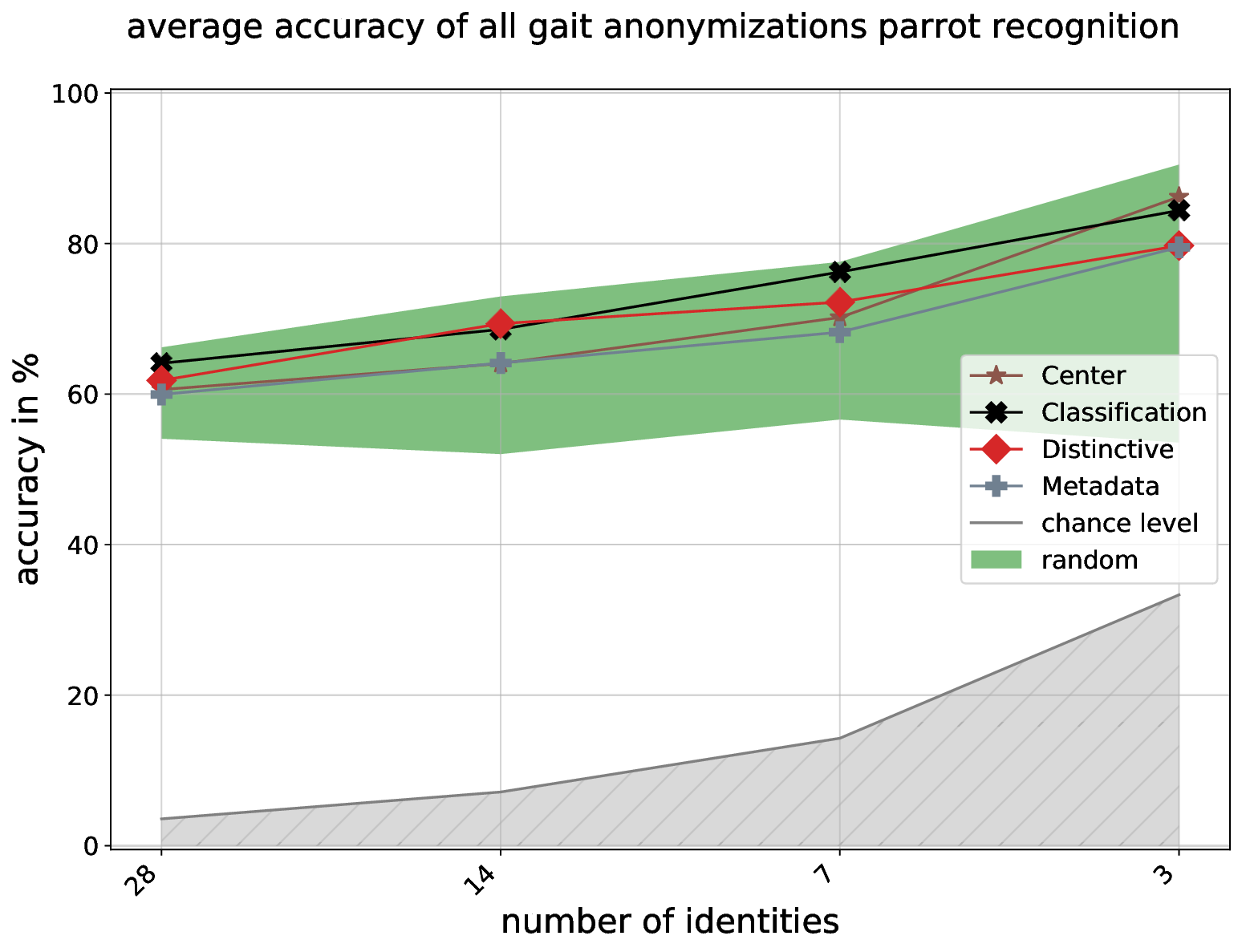}
\caption{All accuracies are given as the average across all gait anonymization techniques. The green area indicates the accuracy range of the previous ten random selections of identities. SVM+flatten is used with parrot recognition.  A lower accuracy means better privacy protection.}\label{img:exp4_gait}
\end{figure}

We present the results of our experiments for H5 for the anonymization of face data in Fig.~\ref{img:exp4_face} and for the anonymization of gait data in Fig.~\ref{img:exp4_gait}.

 Our selection strategies compare \new{to} random selection as follows: our strategies outperform when the number of identities is greater than 62, and under 62 \new{Metadata starts performing worse than the best random selections, while the remaining techniques continue outperforming the best random selections down to 3 identities}. Our \new{Center and Classification} strategies perform best across all numbers of identities, even matching the performance of random selection for 3 identities. What is more, for \new{500 to 15} identities, our Center and Classification strategies \new{increases} over 10\% in performance compared to the best random selection.

For the gait data (Fig.~\ref{img:exp4_gait}), our results \new{are not as good} as for the face data. \new{In general, we find that none of our selection strategies outperforms the best random selections. The strategy that performs consistently best is Classification. It always scores close to the best random selections. The strategy Metadata performs worst, as it does too in the face results. The strategies Center and Distinctive show varying results for different numbers of identities.} \new{Our explanation for the} contrast between face and gait \new{ runs as follows: It is probable that} the significant difference between the number of identities in the full face dataset (n = 1,000) and the number in the full gait dataset (n = 57) \new{results} in less identities to pick from.


The accuracy we achieve with our \new{Classification} selection strategy deserves further attention here, \new{because it performs best across anonymizations for both face and gait}. We will examine \new{Classification} more closely by comparing it to the initial results for our decreases in numbers of identities.

\begin{figure}[!ht]
\centering
      \includegraphics[width=0.47\textwidth]{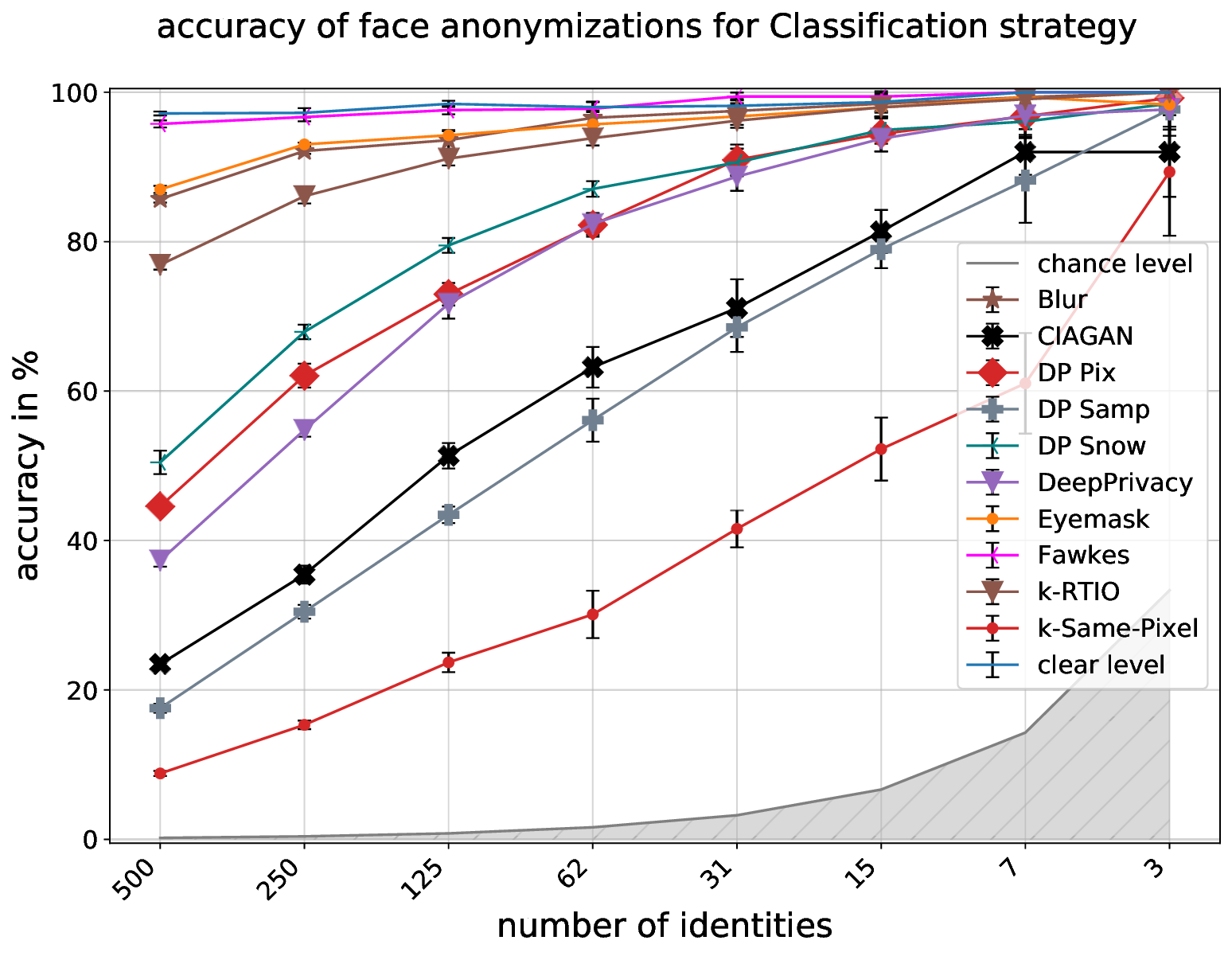}
\caption{
Accuracy of face anonymizations across decreasing numbers of identities. The strategy \new{Center} was used to select the identities. \new{The error bars give the standard deviation over 10 test-train-splits}. ArcFace Retrained is used with parrot recognition on the CelebA dataset. A lower accuracy means better privacy protection.}\label{img:exp4_Classification_face}
\end{figure}

For the face data (see Fig.~\ref{img:exp4_Classification_face}), we observe that clear and Fawkes \new{reach an early plateau close to 100\%} and that Eyemask, Blur, and k-RTIO begin scoring near the 80\% mark and not near the 60\% mark. For 125 identities, Eyemask, Blur, and k-RTIO also plateau earlier. DP Samp increases in accuracy steadily from 500 identities to 31 identities, and from there DP Samp accelerates in performance ultimately to achieve 100\% at 3 identities. \textit{k}-Same-Pixel \new{achieves the lowest accuracies compared to the other anonymization techniques. However, \textit{k}-Same-Pixel follows the same trend as the other techniques by steadily increasing as the identities decrease in number. \new{When we compare} to the random selection (see Fig.~\ref{img:exp2_face}), we see an increase from around 60\% to 90\% for 3 identities. Similar increases can also be found for the other anonymization techniques. We conclude that the Classification strategy is effective in selecting identities that are hard for the anonymization techniques to anonymize.}



\begin{figure}[!ht]
\centering
      \includegraphics[width=0.47\textwidth]{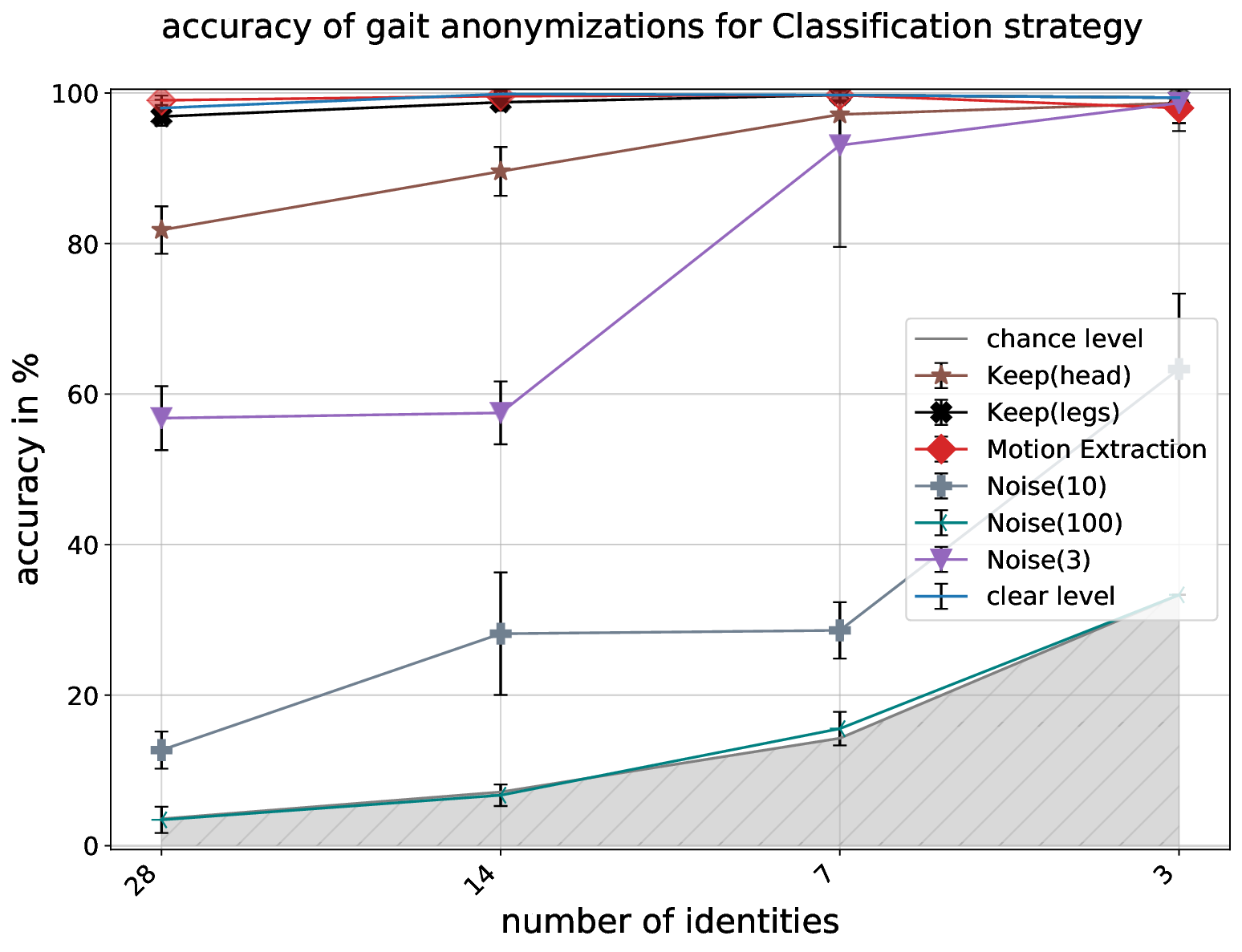}
\caption{Accuracy of gait anonymizations across decreasing numbers of identities. The strategy \new{Classification} was used to select the identities. \new{The error bars give the standard deviation over 10 test-train-splits}. SVM+flatten is used with parrot recognition. A lower accuracy means better privacy protection.}\label{img:exp4_Classification_gait}
\end{figure}

For the gait data (see Fig.~\ref{img:exp4_Classification_gait}), we again find results similar to face. All anonymizations, except Noise(100), score higher. We consider this to be additional evidence that our \new{Classification} strategy is highly successful. \new{Furthermore, we find that the Noise(100) results show that anonymization techniques exist which can achieve near perfect anonymization even in this challenging scenario.}



\subsection{Summary of Results}

\begin{itemize}
\item\new{Recognition performance increases when the system is train\-ed or especially pre-trained on anonymized data.}
\item\new{Recognition performance increases when a reduction is made in the number of identities in the evaluation dataset.}
\item\new{Our Classification selection strategy provides reliable evaluation of anonymization. When, however, the number of identities in the evaluation dataset is very small, Classification might be outperformed by best-case random selections.}
\item\new{Anonymization techniques perform differently across recognition systems. As a direct consequence, it remains unclear which anonymization technique performs best in conjunction with which recognition system.}
\item\new{For some anonymization techniques which use noise injection, it is crucial to determine the optimal proportion of anonymized data for both training and pre-training.}

\end{itemize}

\section{Discussion, Limitations, and Future Work}\label{sec:future_work}

The results of our three experiments confirm all five of our hypotheses. We see ourselves justified in drawing the overall conclusion that our \new{methodological }recommendation\new{s} will improve the state-of-the-art in \new{the evaluation of the anonymization of biometric data}.

Our results for \new{the Hypotheses }H1 and H2 clearly show that training and also pre-training with anonymized data significantly improves the performance of the recognition and thus opens the door to improved evaluation of face and gait anonymization. As demonstrated for face anonymization, even a small amount of anonymized data greatly improves the training process. However, our results also indicate that an excess of noisy training data may decrease the performance. Therefore, for anonymization by noise injection (e.g. Laplace mechanism), we conclude that \new{care} should be \new{taken to determine} the right amount of anonymized training data. Nonetheless, we draw the final conclusion that training with anonymized data significantly improves the validity of the evaluation methodology. Without anonymized data in the train set, the performance of the anonymization is bound to be overestimated.

Our results for \new{the Hypothesis} H3 show that the recognition systems which perform comparably to one another on clear data may perform differently \new{from one another }on anonymized data. Since the performance on clear data is not a good predictor of performance on anonymized data, we conclude that the recognition system which seems to perform at the state-of-the-art on clear data \new{might} not accurately evaluate anonymization performance. \new{This} holds especially for anonymizations which use noise injection, as demonstrated by our results for gait anonymization. Therefore, we consider it the minimum that multiple recognition systems be used with different model architectures. \new{Furthermore, we recommend designing recognition systems to be more resistant to anonymization. Our reason is clear: there is no single recognition system that performs best in all cases, not for face anonymization and not for gait anonymization. Understanding which recognition system architecture works best for which anonymization together with training the system on anonymized data will help to achieve more reliable evaluation results.}

Our results for \new{the Hypothesis} H4 confirm that for most anonymization techniques, a reduced number of identities in the evaluation dataset increases the recognition performance\new{ more than what the increase in chance-level can explain.} This reduction in the number of identities presents a more challenging scenario for the anonymization. Our results for H4 also show that, as the number of identities decreases, the run-to-run variation of possible results increases. We conclude that the selection of identities for the subset is a significant task in the evaluation of anonymization performance. 

Our results for \new{the Hypothesis} H5 \new{clearly indicate} that a more challenging dataset is generated when our \new{Classification} selection strategy is used to select the identities for a reduced evaluation dataset. \new{However, it appears that for very small datasets, multiple random selections can still outperform our Classification selection strategy. Hence, we recommend performing Classification and additionally the random selections in order to determine which performs best at identity selection for the evaluation dataset.}




All in all, \new{our} proposed improvements \new{will} evaluate biometric anonymization techniques \new{much more convincingly than these techniques are currently being evaluated}. Further research, however, is clearly necessary. For example, \new{our methodological} improvements \new{will need to be validated} on other biometric traits.
\new{In addition, it remains an open research question precisely} which types of recognition systems perform best on which types of anonymization. Answers here will help decide whether, in fact, a systematic approach exists for building recognition systems that perform well on specific anonymizations.

\section{Conclusion}\label{sec:conclusion}
Biometric recognition technologies\new{,} such as face recognition systems\new{,} pose a real threat to privacy. Therefore, \new{a crucial technique for} privacy protection \new{is anonymization}, and likewise,  evaluation is crucial to anonymization. This paper assesses the state-of-the-art methodologies used for the evaluation of anonymization techniques, \new{finds} flaws \new{in those methodologies}, and \new{proposes how the methodologies can be improved}.

We find several major flaws in the state-of-the-art methodologies for the evaluation of biometric anonymization. The state-of-the-art evaluation is based on weak and unrealistic assumptions about the adversary. The\new{se} adversaries act in ignorance of the anonymization in place and are accordingly unable to adapt their recognition systems. These are not real\new{istic} adversaries of anonymization techniques. Therefore, the state-of-the-art methodologies largely fail to assess accurately the performance of the recognition. 



To begin the work of correcting such flaws, we have proposed to improve the evaluation methodology for the anonymization of biometric data. \new{It is our recommendation} that recognition systems \new{which are }trained not only \new{on} clear data but also \new{on} anonymized data be used to evaluate anonymization performance. Furthermore, we argue that the use of a variety of different recognition systems will improve the \new{rigor} of the evaluation. The use of merely a single classifier trained \new{only }on clear data \new{might result in} unreliable, overoptimistic estimates of anonymization performance. Hence, we recommend using multiple recognition systems trained on anonymized data.
And lastly, we recommend \new{using} a more challenging evaluation dataset to approximate worst-case performance. Our results indicate that such a dataset can be constructed by reducing the number of identities and selecting the easy-to-distinguish identities with our proposed \new{Classification} strategy. \new{We have proposed improvements to the state-of-the-art in evaluation methodologies that will pre-empt overestimations of biometric anonymization performance. We have backed this finding with strong experimental evidence. Thus, we conclude that our proposed improvements lay the cornerstone of a more reliable evaluation methodology for the anonymization of biometric data.}

\begin{acks}
We would like to thank our in-house textician for his helpful feedback and valuable contributions in editing our manuscript, which have improved the clarity and scientific quality of our research.
This work has been funded by the German Research Foundation (DFG, Deutsche Forschungsgemeinschaft) as part of Germany’s Excellence Strategy – EXC 2050/1 – Project ID 390696704 – Cluster of Excellence “Centre for Tactile Internet with Human-in-the-Loop” (CeTI) of Technische Universität Dresden, and by funding of the Helmholtz Association through the KASTEL Security Research Labs (HGF Topic 46.23).
\end{acks}

\bibliographystyle{ACM-Reference-Format}
\bibliography{Methodology_Survey}

\appendix

\section{Survey of State-of-the-Art Evaluation Methodology for the Anonymization of Biometric Data}\label{sec:state-of-the-art}

\new{In order to learn about the current state-of-the-art for evaluating biometric anonymization, we perform a survey study on 49 papers (see Table~\ref{tab:list_survey_papers}). The majority of papers are from Hanisch et. al.~\cite{hanisch_privacy-protecting_2021}, which collected papers that perform behavioral data anonymization and include traits like voice, gait, and brain activity. Additionally, we use the corpus of face anonymization papers from a survey by Ribaric et. al.~\cite{ribaric_-identification_2016}, which focuses on anonymization in media content. We filtered the papers to match our scenario.}

\new{Our first category for separating the corpus is the \textit{biometric trait} which the anonymization tries to protect and also the \textit{protection goal}. The protection goal may be either to prevent identity disclosure or attribute disclosure.
Since the anonymization approaches are tested against a biometric recognition system, we note whether the evaluations rely on a single approach or test \textit{multiple recognition systems}. 
Further, we examine whether \textit{multiple parameters} for the anonymization technique are evaluated. Our main interest in this survey was to learn which kind of attacker model the evaluations employed. For this, we compare whether an \textit{open-set} or \textit{closed-set} model was applied and with which kind of \textit{training data} (clear or anonymized) the recognition system was trained. Further, we check if the \textit{reversibility} of the anonymization approach was tested. Moreover, we compare the different \textit{metrics} employed to measure the anonymization performance.}

\new{Taking together all the reviewed papers, we find that most focus on anonymizing voice data, then face, gait, and hand. 
Only one paper tackles brain activity and one other tackles eye-gaze (see Table~\ref{tab:trait}). Most papers try to protect against identity inference, while six paper try to protect against attribute inference, and five against both identity inference and attribute inference. 
Regarding metrics for the measurement of privacy protection, we find that accuracy (also including metrics closely based on accuracy e.g. $1 - accuracy$) is the most commonly used metric, followed by the equal error rate (EER). 
Some uncommon metrics we observed were the usage of F1-Score~\cite{matovu_jekyll_2018}, and half total error rates (HTER)~\cite{matovu_your_2016}}.

\new{As seen in Table~\ref{tab:overview} slightly more than half of the papers evaluate different parameter configurations for their anonymization technique, while only about one in four papers uses more than one recognition system for its evaluation. For the recognition scenario, we find that most papers use a closed-set approach. When it comes to training the recognition system, all papers use clear data for training, only a minority also trains the recognition system with anonymized data. For the test whether the anonymization technique can be reversed, we find only one paper~\cite{qian_hidebehind_2018} that considers this for the evaluation, although it only performed a theoretical analysis.} 

\begin{table}
\centering
\begin{tabular}{c|p{5.5cm}}
Trait & Papers (Count and Sources)\\
\hline
Voice & 22 (\cite{abou-zleikha_discriminative_2015},~\cite{bahmaninezhad_convolutional_2018},~\cite{aloufi_emotionless_2019},~\cite{hamm_enhancing_2017},~\cite{lal_srivastava_evaluating_2020},~\cite{pribil_evaluation_2018},~\cite{nelus_gender_2018},~\cite{qian_hidebehind_2018},\newline~\cite{lopezotero_influence_2017},~\cite{parthasarathi_lp_2011},~\cite{keskin_measuring_2019},~\cite{pobar_online_2014},~\cite{nelus_privacy-aware_2019},~\cite{hashimoto_privacy-preserving_2016},~\cite{portelo_privacy-preserving_nodate},~\cite{fang_speaker_2019},\newline~\cite{justin_speaker_2015},~\cite{qian_speech_2021},~\cite{cohen-hadria_voice_2019},~\cite{jin_voice_2009},~\cite{vaidya_you_2019})\\

Face & 10 (\cite{neustaedter_blur_2006},~\cite{meng_face_2014},~\cite{gross_integrating_2005},~\cite{gross_model-based_2006}, ~\cite{newton_preserving_2005},~\cite{gross_face_2009},~\cite{meng_retaining_2014},\newline~\cite{boyle_effects_2000},~\cite{korshunov_using_2013},~\cite{korshunov_using_2013-1})\\

Gait & 8 (\cite{agrawal_person_2011},~\cite{hirose_anonymization_2019},~\cite{ivasic-kos_person_2014}, ~\cite{jourdan_toward_2018},~\cite{matovu_jekyll_2018},~\cite{tieu_rgb_2019},~\cite{tieu_approach_2017},~\cite{tieu_spatio-temporal_2019})\\

Brain Activity & 2 (\cite{yao_improved_2019},~\cite{matovu_your_2016})\\

Eye-gaze & 2 (\cite{steil_privacy-aware_2019},~\cite{bozkir_differential_2020})\\

Hand & 5 (\cite{leinonen_preventing_2017},~\cite{maiorana_bioconvolving_2011},~\cite{migdal_keystroke_2019},~\cite{monaco_obfuscating_2017},~\cite{vassallo_privacy-preserving_2017})\\
\end{tabular}
\caption{Publications included in the state-of-the-art survey with corresponding trait}\label{tab:list_survey_papers}
\end{table}

\begin{table}
\centering
\begin{tabular}{ l | c c c c c c}
Trait& Voice & Face & Gait & Hand & Brain & Eye \\ 
& 22 & 10 & 8 & 5 & 2 & 2 \\  
\hline
Protection Goal & \multicolumn{2}{c}{Identity}  & \multicolumn{2}{c}{Attribute}  & \multicolumn{2}{c}{Both}  \\ 
 & \multicolumn{2}{c}{38} & \multicolumn{2}{c}{6} & \multicolumn{2}{c}{5} \\ 
\hline
Metric & \multicolumn{2}{c}{Accuracy}  & \multicolumn{2}{c}{EER}  & \multicolumn{2}{c}{Other}  \\ 
 & \multicolumn{2}{c}{36} & \multicolumn{2}{c}{10} & \multicolumn{2}{c}{3}\\
\end{tabular}
\caption{Publication count for biometric trait, protection goal, and metric to evaluate the technique}
\label{tab:trait}
\end{table}

\begin{table}
\centering
\begin{tabular}{ l | c c c c c c}
 & \multicolumn{3}{c|}{Yes} & \multicolumn{3}{|c}{No} \\ 
 \hline
anonymized training data & \multicolumn{3}{c|}{8} & \multicolumn{3}{|c}{41}\\  
test reversibility &  \multicolumn{3}{c|}{1} &\multicolumn{3}{|c}{48}\\
closed-set assumption &  \multicolumn{3}{c|}{38} & \multicolumn{3}{|c}{11}\\
multiple parameters &  \multicolumn{3}{c|}{28} & \multicolumn{3}{|c}{21} \\
multiple recognition systems &  \multicolumn{3}{c|}{12} & \multicolumn{3}{|c}{37}\\
\end{tabular}
\caption{The number of papers for the remaining categories}
\label{tab:overview}
\end{table}

\section{Additional Results}\label{sec:additional_results}

\begin{figure}[!ht]
\centering
      \includegraphics[width=0.47\textwidth]{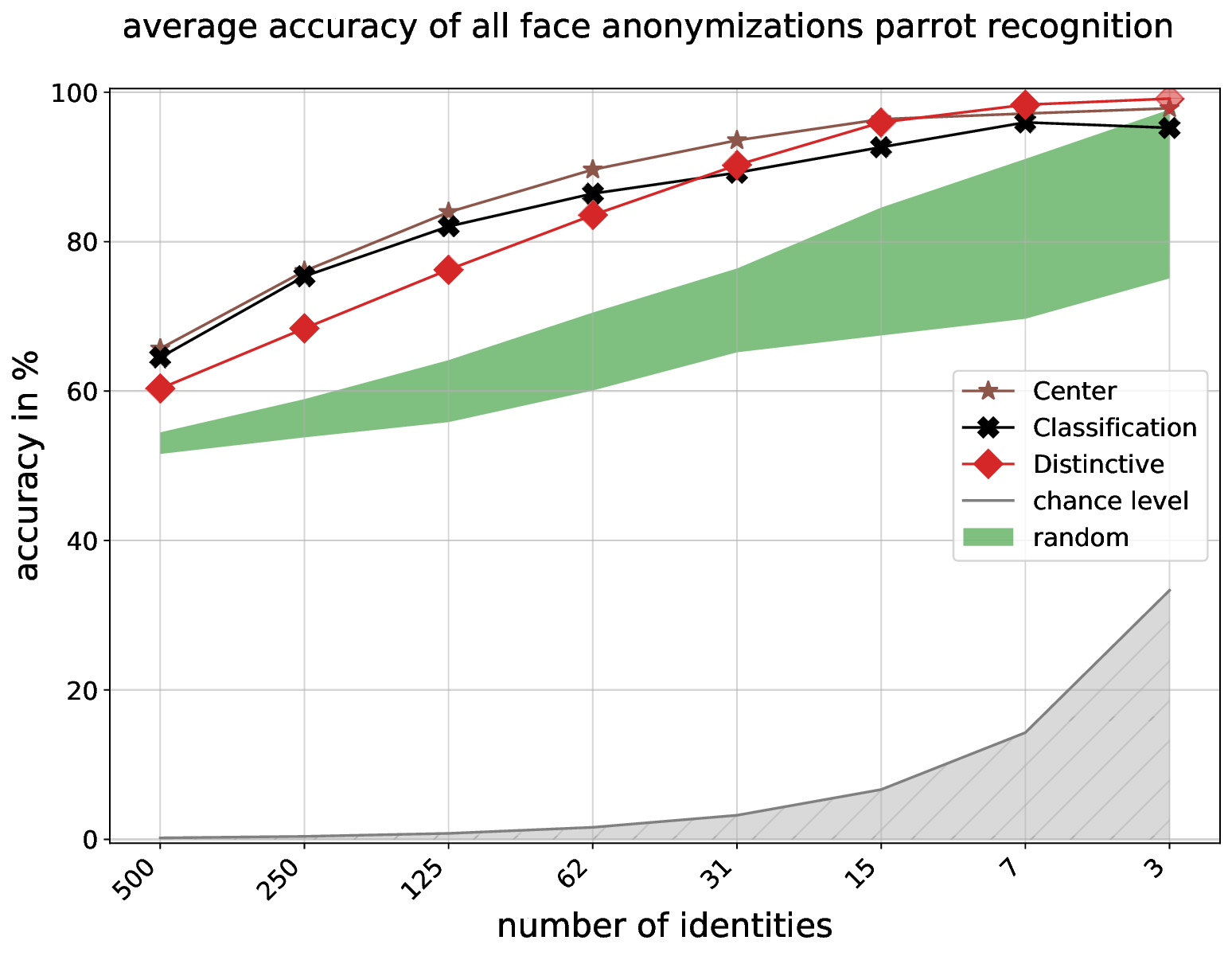}
\caption{All accuracies are given as the average across all face anonymization techniques. The green area indicates the accuracy range of the previous ten random selections of identities. ArcFace Retrained is used with parrot recognition on the WebFace260M dataset. A lower accuracy means better privacy protection.}\label{fig:exp4_All_WebFace}
\end{figure}

\begin{figure}[!ht]
\centering
      \includegraphics[width=0.47\textwidth]{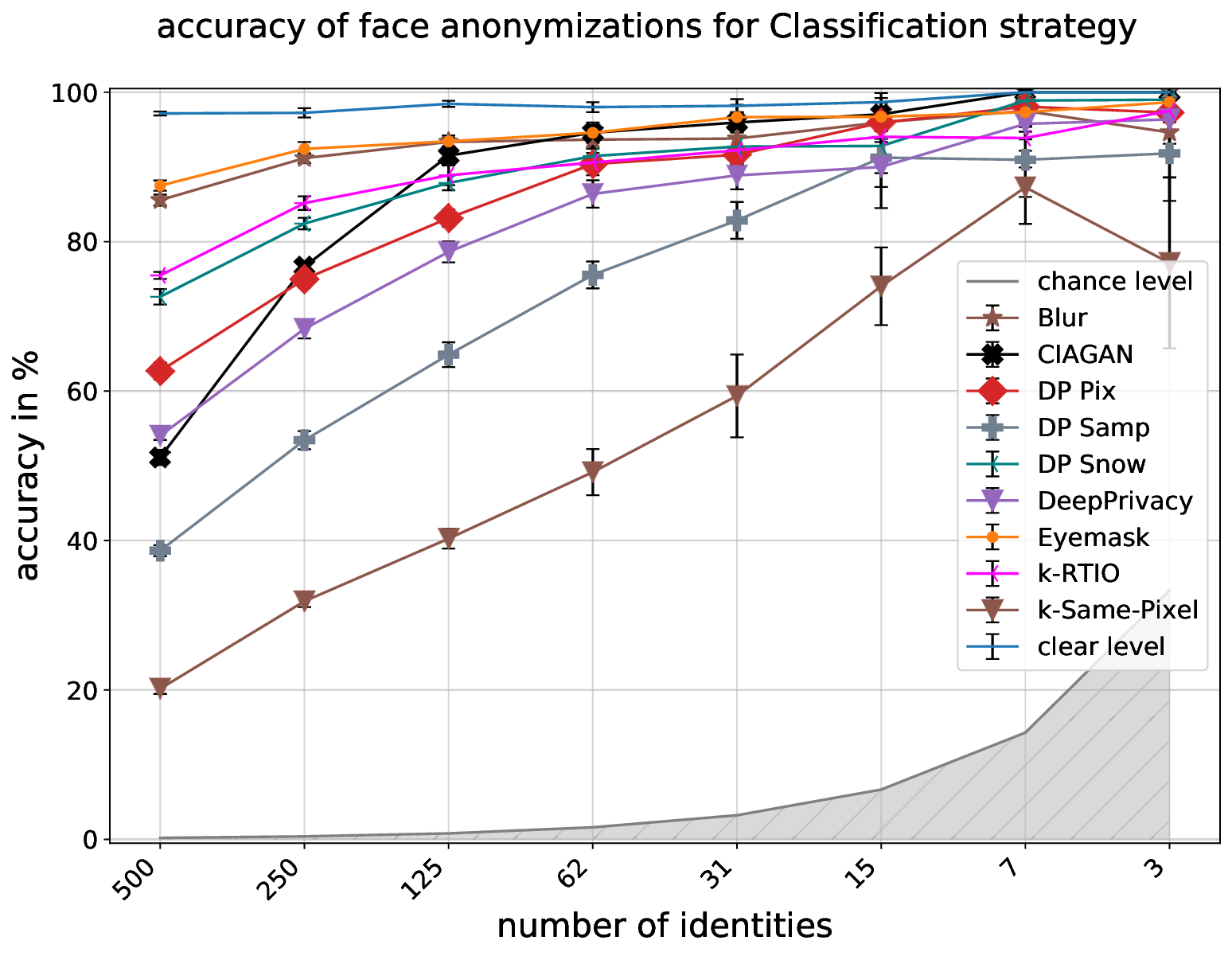}
\caption{
Accuracy of face anonymizations across decreasing numbers of identities. The strategy \new{Center} was used to select the identities. \new{The error bars give the standard deviation over 10 test-train-splits}. ArcFace Retrained is used with parrot recognition on the WebFace260M dataset. A lower accuracy means better privacy protection.}\label{fig:exp4_Classification_WebFace}
\end{figure}

\begin{figure}[!ht]
\centering
      \includegraphics[width=0.47\textwidth]{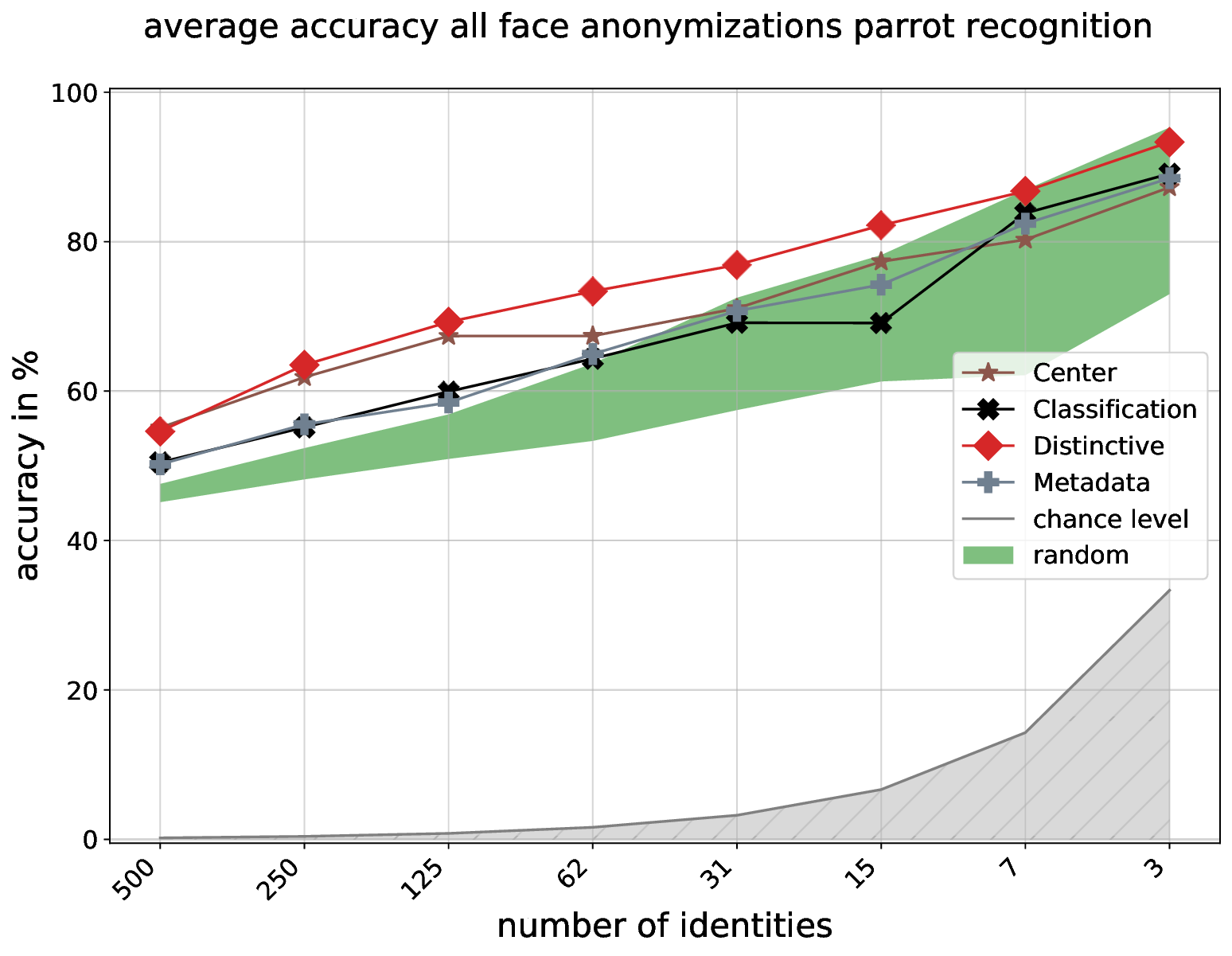}
\caption{All accuracies are given as the average across all face anonymization techniques. The green area indicates the accuracy range of the previous ten random selections of identities. The selection have been performed on clear data. ArcFace Retrained is used with parrot recognition on the CelebA dataset. A lower accuracy means better privacy protection.}\label{fig:exp4_all_select_on_clear_celebA}
\end{figure}

\begin{figure}[!ht]
\centering
      \includegraphics[width=0.47\textwidth]{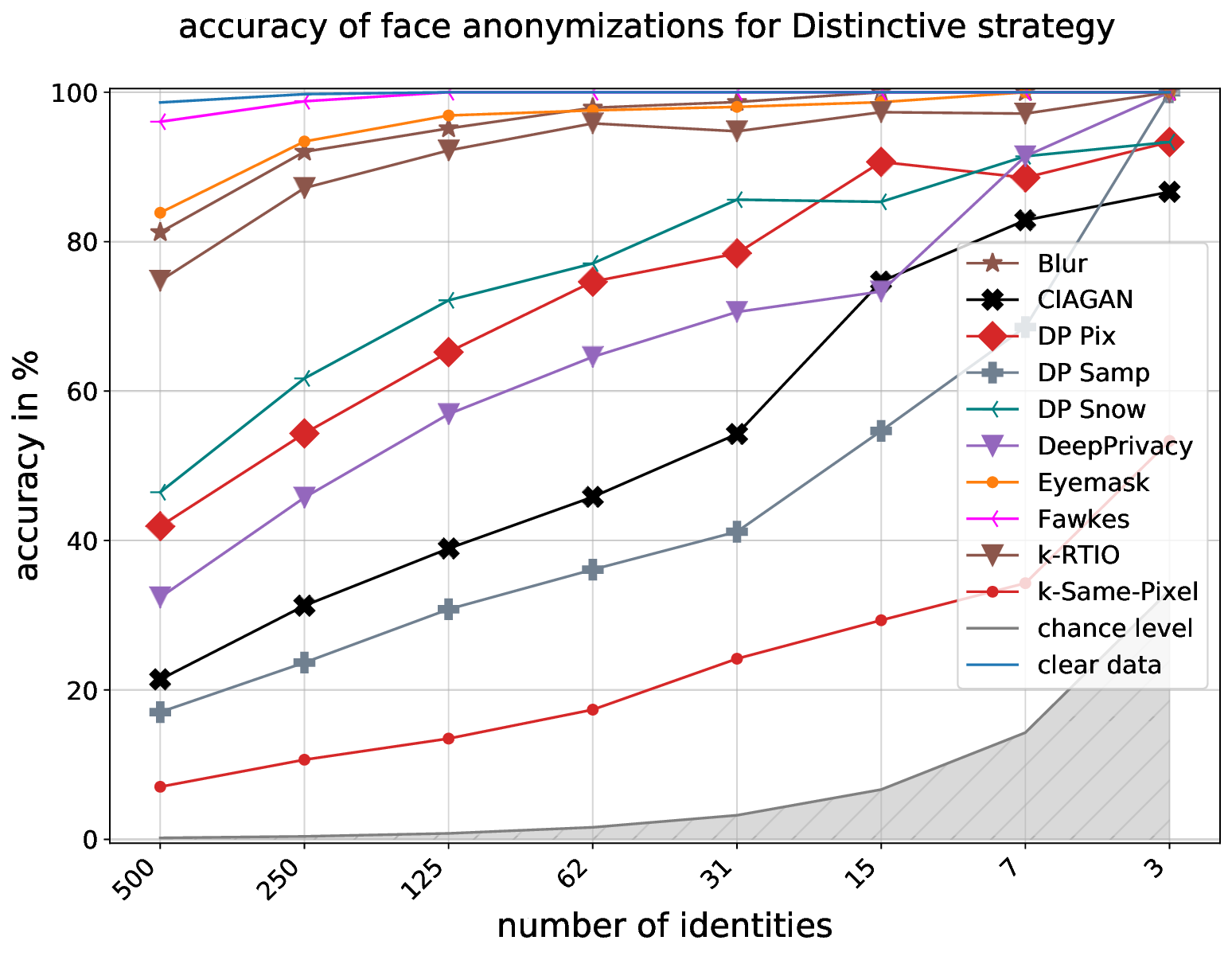}
\caption{
Accuracy of face anonymizations across decreasing numbers of identities. The strategy Distinctive was used to select the identities on clear data. ArcFace Retrained is used with parrot recognition on the CelebA dataset. A lower accuracy means better privacy protection.}\label{fig:exp4_Distinctive_select_on_clear_celeba}
\end{figure}

\new{In the following we report additional results of our experiments, this includes the reproduction of H4 and H5 on the WebFace260M dataset (see Fig.~\ref{fig:exp4_All_WebFace} and Fig.~\ref{fig:exp4_Classification_WebFace}), the Metadata strategy has been excluded as no soft biometric information of the identities was available. We also report the performance of our selection strategies on clear, instead of anonymized, data on the CelebA dataset for all strategies (see Fig.~\ref{fig:exp4_all_select_on_clear_celebA}) and the Distinctive strategy (see Fig.~\ref{fig:exp4_Distinctive_select_on_clear_celeba}) in particular.}

\end{document}